\documentclass[a4paper,aps,prd,twocolumn,tightenlines,preprintnumbers,nofootinbib,showkeys,superscriptaddress]{revtex4-1}

\pdfoutput=1

\usepackage{fullpage}
\usepackage{amsfonts}
\usepackage{amsmath}
\usepackage{slashed}
\usepackage{amssymb}
\usepackage{graphicx}
\usepackage{makeidx}
\usepackage{cancel}
\usepackage{epic}
\usepackage{eepic}
\usepackage{epsfig}
\usepackage{latexsym}
\usepackage[dvipsnames]{xcolor}
\usepackage{float}
\usepackage{multirow}
\usepackage[export]{adjustbox}
\usepackage{xurl,hyperref}
\usepackage{enumitem}
\hypersetup{colorlinks=true,citecolor=red,linkcolor=NavyBlue,urlcolor=NavyBlue}
\usepackage[utf8]{inputenc}
\usepackage[caption=false]{subfig}
 
\usepackage{natbib}
\usepackage{relsize}
\usepackage[left=1.5cm,right=1.4cm,top=2.0cm,bottom=2.cm]{geometry}
\usepackage{mathptmx}
\begin{document}
\relscale{1.05}
\captionsetup[subfigure]{labelformat=empty}

\title{Enhancing Scalar Productions with Leptoquarks at the LHC}

\author{Arvind Bhaskar}
\email{arvind.bhaskar@research.iiit.ac.in}
\affiliation{Center for Computational Natural Sciences and Bioinformatics, International Institute of Information Technology, Hyderabad 500 032, India}

\author{Debottam Das}
\email{debottam@iopb.res.in}
\affiliation{Institute of Physics, Sachivalaya Marg, Bhubaneswar 751 005, India}
\affiliation{Homi Bhabha National Institute, Training School Complex, Anushakti Nagar, Mumbai 400 085, India}

\author{Bibhabasu De}
\email{bibhabasu.d@iopb.res.in}
\affiliation{Institute of Physics, Sachivalaya Marg, Bhubaneswar 751 005, India}
\affiliation{Homi Bhabha National Institute, Training School Complex, Anushakti Nagar, Mumbai 400 085, India}

\author{Subhadip Mitra}
\email{subhadip.mitra@iiit.ac.in}
\affiliation{Center for Computational Natural Sciences and Bioinformatics, International Institute of Information Technology, Hyderabad 500 032, India}

\date{\today}
\preprint{IP-BBSR/2020-2}

\begin{abstract}
\noindent
The Standard Model (SM), when extended with a leptoquark (LQ) and right-handed neutrinos, can have interesting new implications for Higgs physics. 
We show that sterile neutrinos can induce a boost to the down-type quark Yukawa interactions through a diagonal coupling associated with the quarks and a scalar LQ of electromagnetic charge $1/3$. The relative change is moderately larger in the case of the first two generations of quarks, as they have vanishingly small Yukawa couplings in the SM. The enhancement in the couplings would also lead to a non-negligible contribution from the quark fusion process to the production of the 125 GeV Higgs scalar in the SM, though the gluon fusion always dominates. However, this may not be true for a 
general scalar. As an example, we consider a scenario with a SM-gauge-singlet scalar $\phi$ where an $\mathcal O(1)$ coupling between $\phi$ and the LQ is allowed. The $\phi q \bar{q}$ Yukawa couplings can be generated radiatively only through a loop of LQ and sterile neutrinos. Here, the quark fusion process can have a significant cross section, especially for a light $\phi$. It can even supersede the normally dominant gluon fusion process for a 
moderate to large value of the LQ mass. This model can be tested/constrained at the high luminosity run of the LHC through a potentially large branching fraction of the scalar to two jets.
\end{abstract}
	
\maketitle
	
\section{Introduction}
\noindent	
The discovery of a Standard Model--(SM-) like Higgs boson of mass $125$ GeV at the 
LHC~\cite{Aad:2012tfa,Chatrchyan:2012xdj} and the subsequent measurements of its couplings to other SM particles have 
played a significant role in understanding the possible physics beyond the Standard Model (BSM). The Higgs couplings to the 
third generation fermions and the vector bosons have already been measured within $10$\%--$20$\% of their SM 
predictions~\cite{Aad:2019mbh}. However, it is difficult to put strong bounds on the Yukawa couplings ($y_f$) of the first two 
generations of fermions. 
This is an interesting point since, at the LHC, a change in the
light-quark Yukawa couplings opens up the possibility of light quarks contributions to the production of a Higgs. It motivates us to investigate whether it is possible to enhance the Yukawa couplings of the first two generation quarks in some existing minimal extension of the SM.

Therefore, in this paper, we study a simple extension of the SM augmented with a scalar leptoquark (LQ) of electromagnetic charge $1/3$ (generally denoted as $S_1$) and  right-handed neutrinos. We find that the Yukawa couplings of the down-type quarks receive some new contributions and, for perturbative values of the free coupling parameters, can be moderately enhanced, especially for a SM-like Higgs ($h_{125}$). However, for a singlet Higgs ($\phi$), this enhancement could be more significant and could open up the $q\bar q\to \phi$ production channel.
Here, we systematically study the production of both $h_{125}$ and $\phi$ at the $14$ TeV LHC through the quark and gluon fusion channels in the presence of a $S_1$ and right-handed neutrinos.

LQs are bosons that couple simultaneously to a quark and a lepton. They appear quite naturally in several extensions of the SM, especially in theories of grand unification like the Pati-Salam model~\cite{PhysRevD.8.1240}, $SU(5)$~\cite{Glashow}, or $SO(10)$~\cite{Georgi} (for a review, see \cite{Dorsner:2016wpm}). Though, in principle, LQs can be either scalar or vector in local quantum field theories, the scalar states are more attractive, as the vector ones may lead to some problems with loops~\cite{Blumlein:1996qp,Fajfer:2015ycq,Barbieri:2015yvd}. 
In recent times, LQ models (with or without  right-handed neutrinos) have drawn attention for various reasons. For example, they can be used to explain different
B-meson anomalies~\cite{Dorsner:2013tla,Gripaios:2014tna,Becirevic:2015asa,Becirevic:2016yqi,Crivellin:2017zlb,Cline:2017aed,DiLuzio:2017chi,Mandal:2018kau,Aydemir:2019ynb,Crivellin:2019dwb} 
 or to enhance flavor violating decays of Higgs and leptons like $\tau\rightarrow\mu\gamma$ and $h\rightarrow\tau\mu$~\cite{PhysRevD.93.015010}. 
LQs may also play a role to accommodate dark matter abundance \cite{Mandal:2018czf,Choi:2018stw} or to mitigate the discrepancy in the anomalous magnetic moment of muon 
$(g-2)_\mu$~\cite{Djouadi1990,Cheung:2001ip,Dorsner:2019itg}.
Direct production of TeV scale  right-handed neutrinos at the LHC can be
strongly enhanced if one considers the neutrino mass generated at tree level via the inverse-Seesaw mechanism within LQ scenarios~\cite{Das:2017kkm}. An $S_1$-Higgs coupling can help to stabilize the electroweak vacuum~\cite{Bandyopadhyay:2016oif}.
The collider phenomenology of various LQs has also been extensively discussed in the literature~\cite{Dorsner:2016wpm,Mandal:2015lca,Dorsner:2017ufx,Bandyopadhyay:2018syt,Hiller:2018wbv,Biswas:2018iak,Faber:2018afz,Alves:2018krf,Chandak:2019iwj,Padhan:2019dcp,Allanach:2019zfr}.

In the scenario that we consider, there are three generations of right chiral neutrinos in addition to the $S_1$. Generically, such a scenario is not very difficult to realize within the grand unified frameworks. In fact, considering sterile neutrinos in this context is rather well motivated because of the existence of nonzero
neutrino masses and mixings, which have been firmly established by now. It is known that
an $\mathcal{O}(1)$ Yukawa coupling between the chiral neutrinos
and TeV scale masses for the  right-handed 
neutrinos can explain the experimental observations related to neutrino masses and mixing angles even at tree level if one extends SM to a simple setup 
 like the inverse seesaw mechanism \cite{PhysRevLett.56.561,PhysRevLett.56.564,PhysRevD.34.1642} (ISSM). Of course, this requires the presence of an additional singlet neutrino state $X$ in the model.\footnote{ISSM or inverse seesaw extended supersymmetric models may lead to interesting phenomenology at low energy \cite{Deppisch:2004fa,Abada:2011hm,Abada:2012cq,Mondal:2012jv,BhupalDev:2012ru,Banerjee:2013fga,Abada:2014kba,Arganda:2014dta,Arganda:2015naa}}

Interestingly, the production cross sections 
of sterile neutrinos at the LHC can be enhanced significantly if the ISSM is embedded in a LQ scenario~\cite{Das:2017kkm}.
Similarly, as indicated earlier, a $\nu_R$ state in a loop accompanied with $S_1$ may influence the production of a Higgs at the LHC
and its decays to the SM fermions, especially to the light ones. 
Observable effects can be seen in scenarios with a general scalar sector that may 
include 
additional Higgs states, 
 a TeV scale $\nu_R$, and an ${\mathcal{O}(1)}$ Yukawa 
couplings between the left and right chiral neutrinos. In this paper, we shall
explore this in some detail. Notably, the gluon fusion process (ggF) for producing a 
Higgs scalar gets boosted in presence of a LQ \cite{Agrawal:1999bk}.
Our study is general--it can be applied to both the SM-like and BSM Higgs bosons. Specifically, we consider two cases:\\

\noindent
{\bf\underline{A 125 GeV SM-like Higgs boson ($h_{125}$):}} We investigate how the light-quark Yukawa couplings can get some positive boosts. However, obtaining
a free rise of the Yukawa parameters is not possible in our model\footnote{ 
This may be possible in an effective theory with free parameters. For example, Ref.~\cite{Bar-Shalom:2018rjs} considers a dimension-6 operator of the form
$\displaystyle
f_d (H^\dagger H/\Lambda^2)\left(
\bar{q}_LH d_R\right) +{\rm H.c.}$ (where $\Lambda\sim$ TeV) 
in addition to the SM Yukawa terms that contribute differently to the physical quark masses and effective quark Yukawa couplings. Thus, by choosing $f_d$ one may raise the Yukawa parameters while keeping the physical masses unchanged, though this may require some fine-tuning among the parameters of the model. It is important to note that in the presence 
of higher-dimensional operators, a large Yukawa coupling need not induce large correction to the corresponding quark mass always. 

Such enhancements of the light-quark Yukawa couplings can even be probed at the LHC. An analysis of Higgs boson pair production suggests that in the future the High Luminosity LHC (HL-LHC) may offer a handle 
on this~\cite{Alasfar:2019pmn}. An updated analysis, with $3000$ fb$^{-1}$ of integrated luminosity 
suggests (though not in a fully model independent way) that it may be possible to narrow down the $d$- and $s$-quark Yukawa 
couplings to about $260$ and $13$ times to their SM values, respectively~\cite{deBlas:2019rxi}, i.e.,
\begin{align}
 |\kappa_d|\le 260,\qquad|\kappa_s|\le 13,
 \label{eq:ydlimits}
\end{align}
where the Yukawa coupling modifier $\kappa_q$ is defined as
\begin{equation}
\kappa_q=\frac{y_q^{\rm eff}}{y_q^{\rm SM}}.
\label{eq:kappa}
\end{equation}
}
and, as we shall see, for perturbative new couplings and TeV scale new physics masses, the boosts are moderate and lead 
to some enhancement of both production and decays of $h_{125}$ at the LHC. \\

\noindent
{\bf\underline{A singlet scalar $\phi$ (BSM Higgs):}}
We also study the productions and decays of a scalar $\phi$ that is a singlet under the SM gauge group. Such a scalar has been considered 
in different contexts in the literature. For example, it may serve as a dark matter candidate.
Similarly, a singlet scalar can help solve the so-called $\mu$ problem
 in the Minimal Supersymmetric Standard Model~\cite{Ellwanger:2009dp}. 
 To produce such a singlet at the LHC, one generally relies upon its mixing with the doubletlike Higgs states present in the theory. If the mixing is non-negligible, then the leading order production process turns out to be the gluon fusion (though
 vector boson fusion (VBF) may also become relevant in specific cases~\cite{Das:2018fog}). 
 One may also consider the production of $\phi$ through cascade decays of the doublet Higgs state(s). 
 However, such a process is generally much suppressed. 
 Now, as we shall see, in the presence of 
 a scalar LQ and sterile neutrinos we could have a new loop 
 contribution to the quark fusion production process (qqF).
 The LQ would also contribute to the gluon fusion process.
 In such a setup, 
 the singlet Higgs can potentially be tested at the LHC via its decays to the light-quark states.

The rest of the paper is organised as follows. In section \ref{sec:model} we introduce the model Lagrangian and discuss the new interactions. In section \ref{sec:SM_Higgs}, we discuss the additional contributions to the production and decays of $h_{125}$. In section \ref{sec:limit}, we discuss the bounds on the parameters. In section \ref{subsec:analytic_phi}, we investigate the case of the singlet scalar $\phi$. Finally we summarize our results and conclude in section \ref{sec:conclu}.
\section{The Model: A Simple Extension of the SM}\label{sec:model}
\noindent
As explained in the Introduction, our model is a simple extension of the SM with chiral neutrinos and an additional scalar 
LQ of electromagnetic charge $1/3$,
normally denoted as $S_1$. The LQ transforms under the SM gauge group as 
$\displaystyle\left ({\bf\bar{3}},{\bf 1},1/3\right)$ with $Q_{\rm EM}=T_3+Y$. In the notation of 
Ref.~\cite{Dorsner:2016wpm}, the general fermionic interaction Lagrangian for $S_1$ can be written as
	\begin{align}
\nonumber	\mathcal{L}_F = &\ (y_1^{LL})_{ij} (\bar{Q}_L^{Cia}\epsilon^{ab}L_L^{jb})S_1+ (y_1^{RR})_{ij} 
(\bar{u}_R^{Ci}e_R^j)S_1\\&\ + (y_1^{\overline{RR}})_{ij} (\bar{d}_R^{Ci}\nu_R^j)S_1+{\rm H.c.},
\label{eq:lagrangianF}
\end{align}
where we have suppressed the color indices. The superscript $C$ denotes charge conjugation; $\{i,j\}$ and 
$\{a,b\}$ are flavor and $SU (2)$ indices, respectively. The SM quark and lepton doublets are denoted by 
$Q_L$ and $L_L$, respectively. 
We now add the scalar interaction terms to the Lagrangian in Eq.~\eqref{eq:lagrangianF},
	\begin{align}
\mathcal{L}\supset\ \mathcal L_F&\ +\lambda \left(H^\dagger H\right) \left(S^\dag_1S_1\right)+\lambda^\prime\phi 
\left(S^\dag_1S_1\right)\nonumber\\
	&\ +\mu (H^\dagger H)\phi^2 + \frac{1}{2}M^2_\phi \phi^2 + \bar M^{2}_{S_1}\left(S^\dag_1S_1\right).
	\label{eq:lagrangian}
	\end{align}
Here, $H$ denotes the SM Higgs doublet, and $M_\phi$ and $\bar M_{S_1}$ define the bare mass
parameters for $\phi$ and $S_1$, respectively. We denote the physical Higgs field after the electroweak symmetry breaking as $h\equiv h_{125}$. The singlet $\phi$
does not acquire any vacuum expectation value (VEV). Physical masses can be obtained via
\begin{equation}
H = 
\frac{1}{\sqrt{2}}\begin{pmatrix}
0 \\
v + h
\end{pmatrix} \, , \hspace{0.5 cm}
\phi= \phi, 
\end{equation}
where the SM Higgs VEV $v \simeq 246~$GeV. We assume the
mixing between $H$ and $\phi$, controlled by the dimensionless coupling $\mu$ to be small to ensure that the presence of a singlet 
Higgs does not affect the production and decays of $h_{125}$ significantly via mixing. Notice 
that unlike dimensionless $\lambda$or $\mu$, $\lambda^\prime$ is a dimension-$1$ parameter.
We define the physical mass of $S_1$ to be $M_{S_1}$ as
\begin{equation}
M^2_{S_1}~=~\bar M^{2}_{S_1}+ \frac{1}{2}\lambda v^2~.
\end{equation}

The above Lagrangian simplifies a bit if we 
ignore the mixing
among quarks and neutrinos (i.e., set ${\bf V}_{\rm CKM}={\bf U}_{\rm PMNS}=\mathbb{I}$). For example, we can 
expand Eq.~\eqref{eq:lagrangian} for the first generation as
	\begin{align}\label{eq:Lag}
	\mathcal{L}\ \supset&\ \Big\{y_{1}^{LL}\left(-\bar{d}_L^C\nu_L+ \bar{u}_L^C e_L\right)S_1
	+ y_{1}^{RR}\ \bar{u}_R^C e_RS_1\nonumber\\
	&\ + y_{1}^{\overline{RR}}\ \bar{d}_R^C \nu_R S_1 + H.c.\Big\}+\lambda v h \left(S^\dag_1S_1\right)\nonumber\\
	&\ +\lambda^\prime\phi 
\left(S^\dag_1S_1\right) + \frac{1}{2}M^2_\phi \phi^2 + M^{2}_{S_1}\left(S^\dag_1S_1\right),
	\end{align}
where we have simplified $\displaystyle \left(y_1^{X}\right)_{ii}$ as $y_i^{X}$. 
Since the flavor of the neutrino is irrelevant for the LHC, from here on we shall simply write $\nu$ to denote neutrinos.
	\begin{figure*}[t]
	\begin{center}
	 \quad\quad
	 \subfloat[\quad\quad\quad(a)]{\includegraphics[width=0.255\textwidth,valign=c]{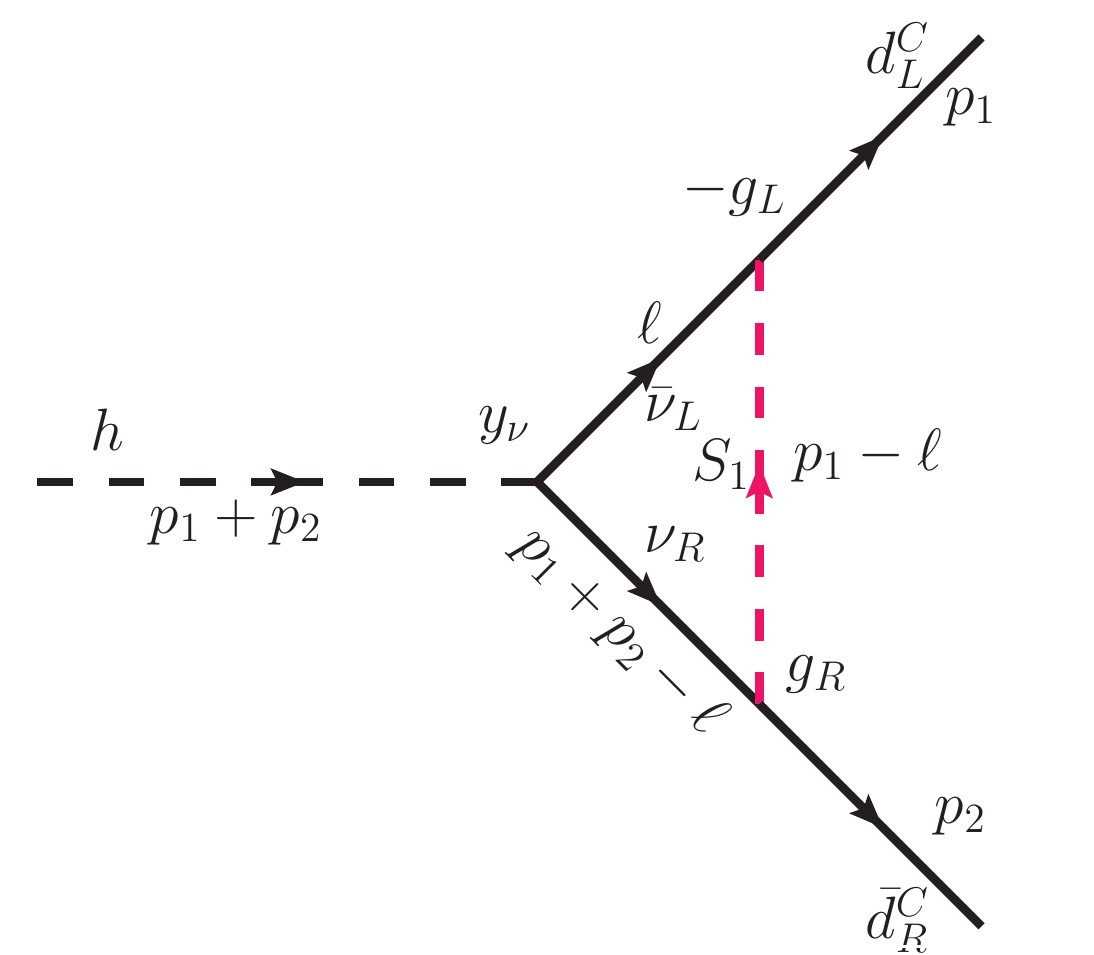}\label{fig:leptoa}} 
	 \subfloat[\quad\quad\quad (b)]{\includegraphics[width=0.255\textwidth,valign=c]{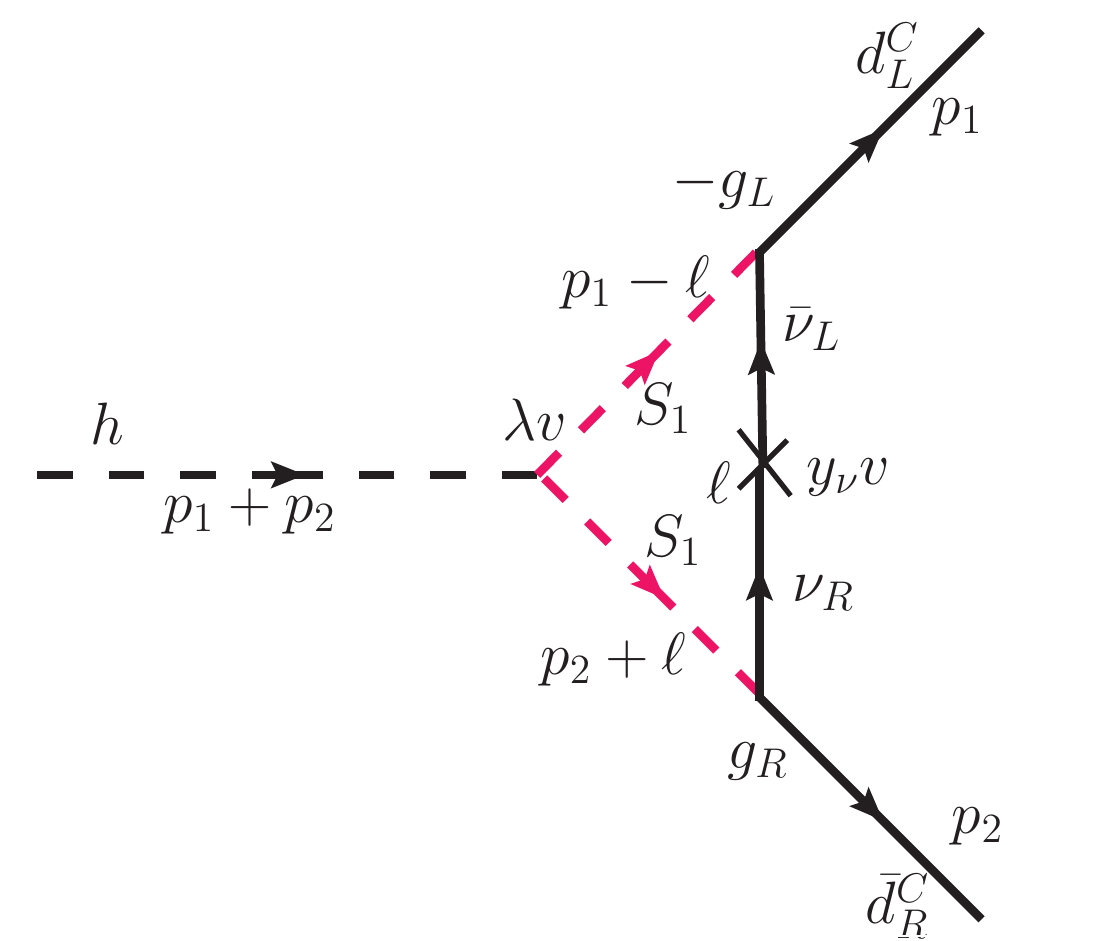}\label{fig:leptob}}
	 \quad\quad
	 \subfloat[\quad\quad\quad (c)]{\includegraphics[width=0.255\textwidth,valign=c]{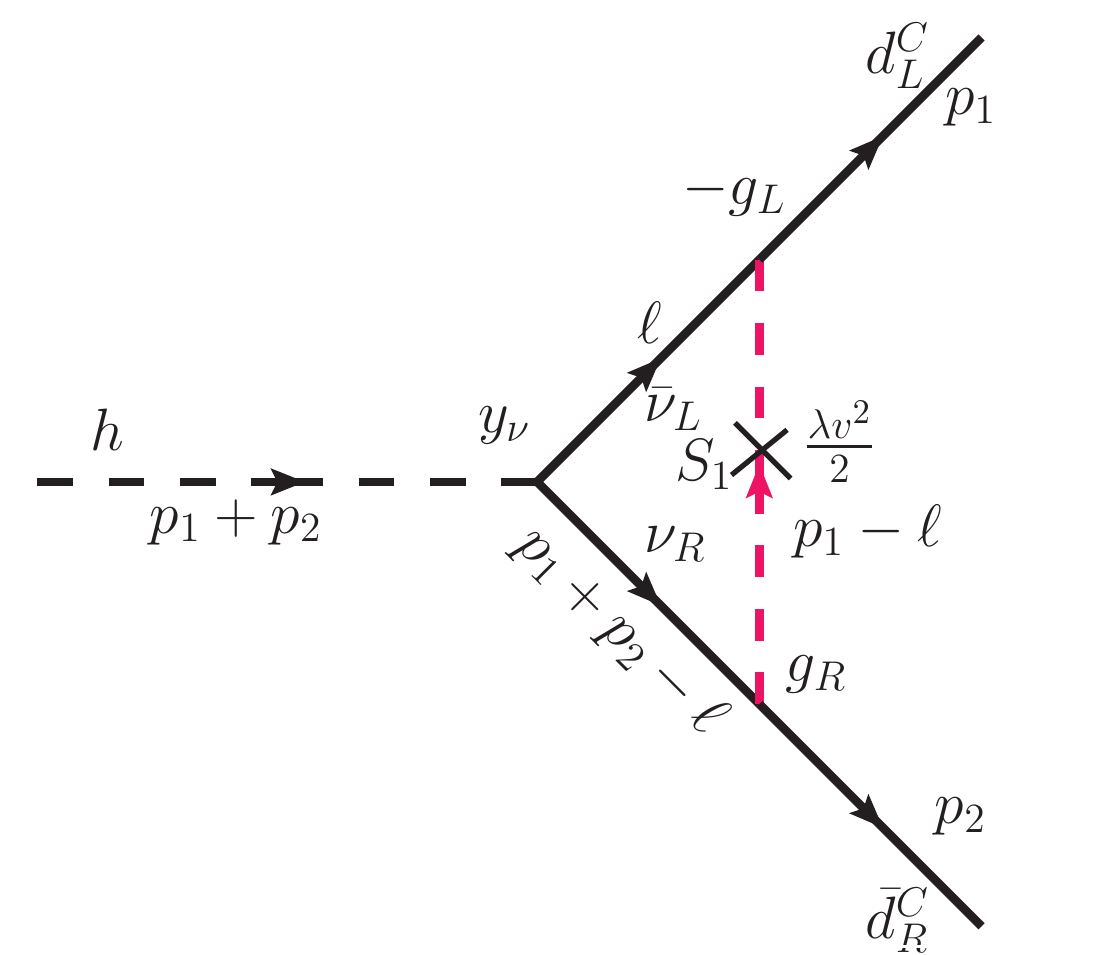}\label{fig:leptoc}}\\
	 	 	\subfloat[\quad\quad\quad (d)]{\includegraphics[width=0.255\textwidth,valign=c]{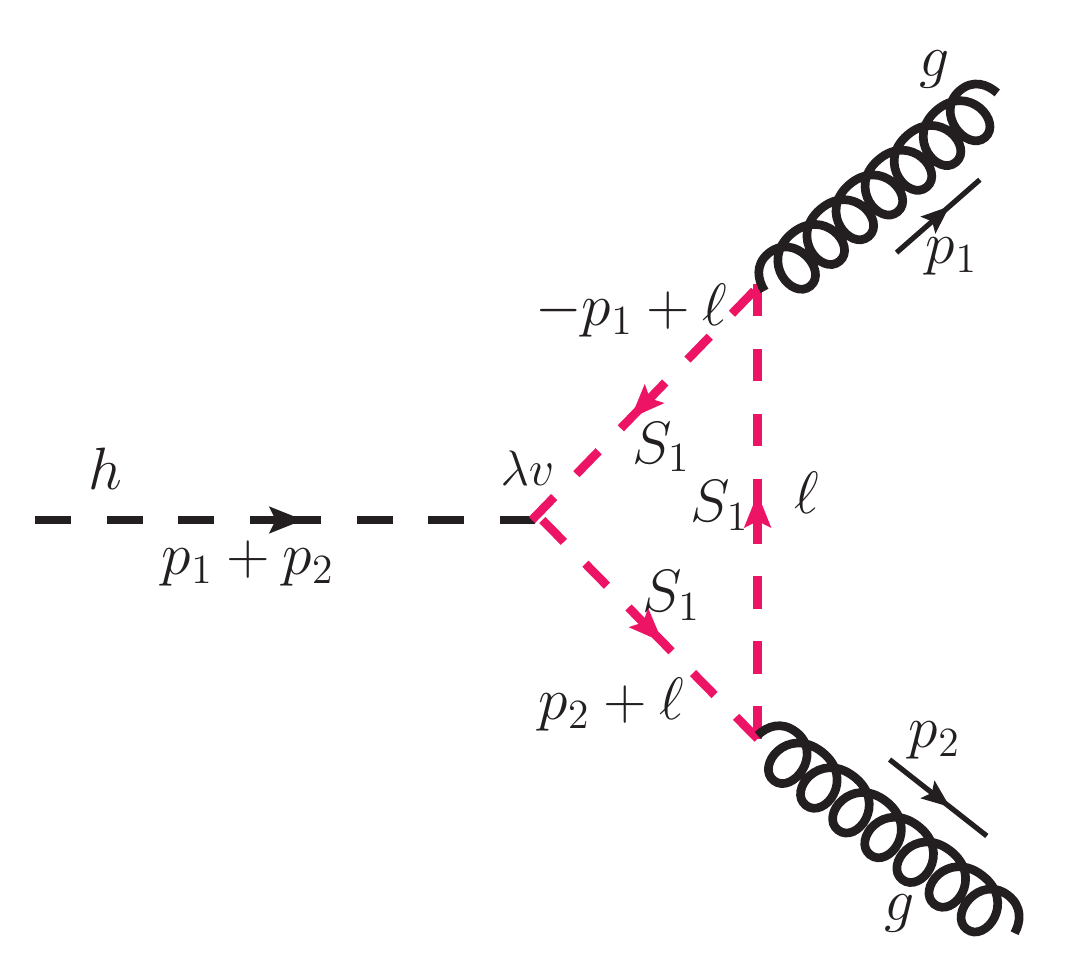}\label{fig:glu_a}}
	 	\quad\quad
	 \subfloat[\quad\quad\quad (e)]{\includegraphics[scale=0.4,valign=c]{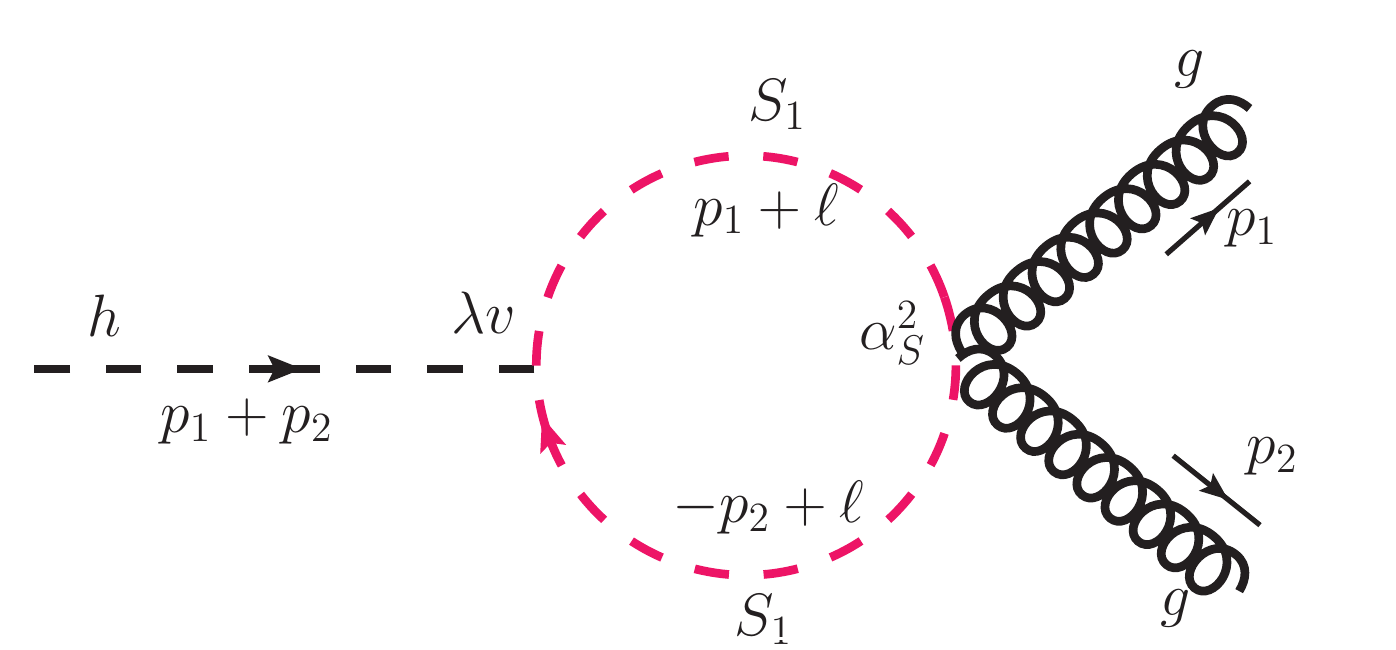}\label{fig:glu_b}
	 \vphantom{\includegraphics[width=0.255\textwidth,valign=c]{lep_q2}}}
	\end{center}
		\caption{ Feynman diagrams showing the SM-like Higgs ($h_{125}$) decaying to (a)--(c) down quarks and (d), (e) gluon pairs through loop diagrams mediated by $S_1$ and chiral neutrinos. Only in (a) and (c) does the Higgs couple to $\nu$ whereas it couples to $S_1$ in all the other diagrams. The couplings $g_L=y_1^{LL}$ and $g_R=y_1^{\overline{RR}}$ [see Eq.~\eqref{eq:Lag}]. The diagrams for $s$- and $b$-quarks are similar to the last two diagrams. Note that we absorb a factor of $1/\sqrt2$ in the definition of Yukawa couplings in the mass basis, i.e., we write $y_\nu$ instead of $y_\nu/\sqrt{2}$. 
		}
		\label{fig:lepto}
	\end{figure*}

The terms in Eq.~\eqref{eq:Lag} have the potential to boost up some 
production/decay modes for $h$ and $\phi$. For example, it would lead to an additional contribution to the effective $hgg$ coupling (see Fig.~\ref{fig:lepto})~\cite{Agrawal:1999bk}. Similarly, the decay $h\to d \bar{d}$, which is negligible 
in the SM, may get a boost 
now, as long as some of the new couplings are not negligible. The processes are illustrated in Figs.~\ref{fig:leptoa}--\ref{fig:leptoc}
[the first diagram is independent of $v$, while the other two are of $\mathcal O\left(v^2\right)$]
where the Higgs is shown to be decaying to a $d \bar d$ pair via a triangle loop mediated by $S_1$ and chiral neutrinos. There 
are two possibilities:  the Higgs directly couples either with the chiral neutrinos or the LQ. Since the contributions of these 
diagrams appear as corrections to $y_d$, it is easy to see that the fermion in the loop (i.e., the neutrino) 
has to go through a chirality flip. In this case, the right-handed neutrino from the third term in Eq.~\eqref{eq:lagrangianF} helps in getting a non-zero contribution. 

One can, of course, imagine similar diagrams with charged leptons 
in the loops, contributing to the $h\to u \bar u$ (or any other 
up-type quark-antiquark pair) decay. 
However, the contributions of such diagrams would be small as they are 
suppressed by the tiny charged lepton Yukawa 
couplings, at least for the first two generations. 
If we restrict ourselves only to flavour diagonal
 couplings in Eq.~\eqref{eq:lagrangianF} (i.e., we allow only $i=j$ terms), only the top Yukawa $y_t$ would be modified appreciably. If we allow off-diagonal couplings, one can get contributions for the first two generations of Yukawa couplings-- namely, $y_u$ and $y_c$, respectively. However, one needs to be careful as off-diagonal LQ-quark-lepton couplings are
constrained, particularly for the first two generations \cite{Dorsner:2016wpm,Mandal:2019gff}. 
In this case, we consider only flavour diagonal couplings and look only at the 
modification of Higgs couplings to down-type quarks. Thus, one may always set 
$\left(y_1^{RR}\right)_{ij}=0$ for all values of $i$ and $j$. 
This may lead to a somewhat favourable situation in some cases to accommodate rare decays of fermions through LQ exchange.

Before we discuss productions and decays of $h_{125}$ and $\phi$ in our model, a few comments are in order.
As we shall see in the next section, an order $1$ 
$h \bar{\nu}_{L}\nu_{R}$ coupling, i.e., $y_\nu \sim \mathcal O(1)$ and a TeV scale 
mass for the $\nu_R$ would be helpful to raise the Yukawa couplings of the light 
quarks. Typically, the models like ISSM would be able to 
accommodate such a scenario. In the ISSM, an additional gauge singlet
neutrino, usually denoted by $X$, is assigned a Majorana mass term $\mu_X XX$ while 
$\nu_R$ receives a Dirac mass term of the form $M \bar {\nu_R} X$. For our purposes, we may assume that this singlet $X$ cannot 
directly interact with any other particle we consider. However, since 
it interacts exclusively with the $\nu_R$ fields via $M$, it would modify the $\nu_R$ propagators. In this case, it may
be useful to define something called a ``fat $\nu_R$ propagator" 
\cite{Arganda:2017vdb} that includes all the effects of the sequential 
insertions of the $X$ field. 
For simplicity, we do not display this interaction and mass term of the  right-handed neutrinos explicitly in Eq.~\eqref{eq:lagrangianF}. 
One can explicitly consider an ISSM in the backdrop of our analysis and easily accommodate fat $\nu_R$ propagators without any change in our results. 

\section{Contribution to the Production and Decays of $\pmb{h_{125}}$}\label{sec:SM_Higgs}
\noindent 
In this section, we first look into the additional contributions to the Yukawa couplings of the down-type quarks with $h_{125}$. The relevant interactions can be read from Eq.~\eqref{eq:Lag}. We shall then discuss the role of these loops in the production of $h_{125}$ and its decays to the down-type quarks. In this paper, 
we compute all the loop diagrams using dimensional regularization and Feynman parametrization and then match the results using the Passarino-Veltman (PV) integrals~\cite{Passarino:1978jh}. We evaluate the PV integrals with two publicly available packages, FeynCalc~\cite{Shtabovenko:2016sxi} and LoopTools~\cite{Hahn:1998yk}.

 \subsection{Correction to Yukawa Couplings of the Down-type Quarks}\label{subsec:analytic_h}
 	\begin{figure}[t]
	\begin{center}
\subfloat[(a)]{\includegraphics[width=0.7\columnwidth]{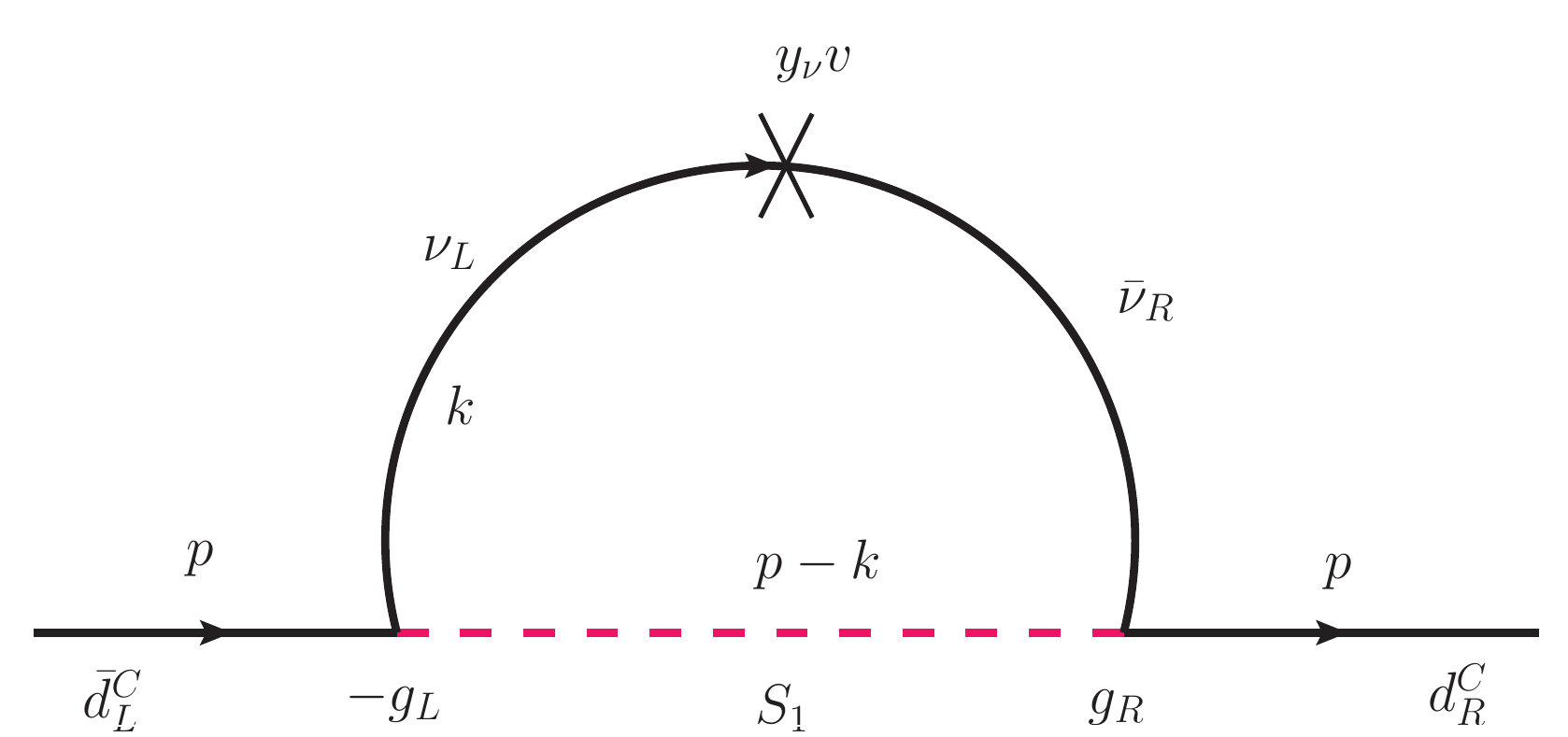}\label{fig:mass1a}}\\
	\subfloat[(b)]{\includegraphics[width=0.7\columnwidth]{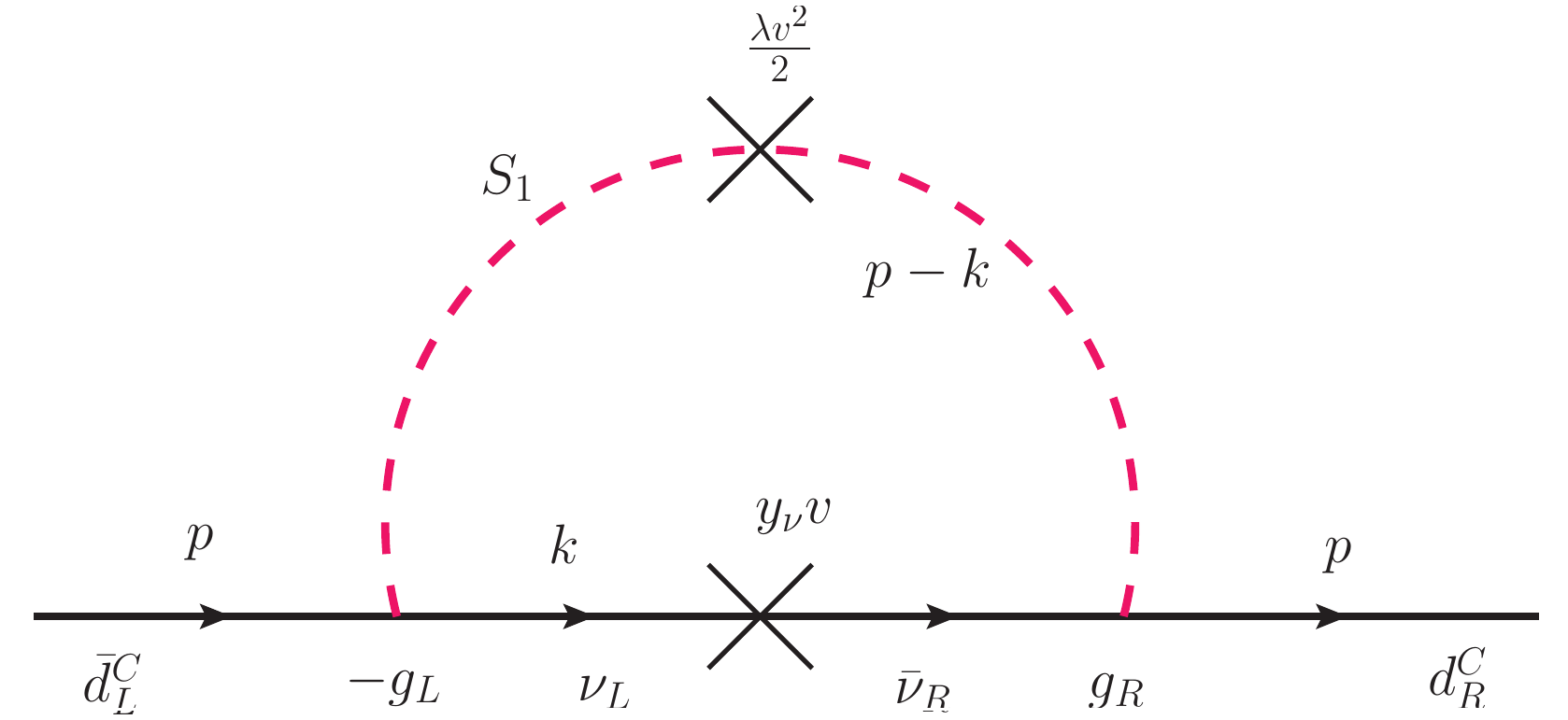}\label{fig:mass1b}}
\caption{Feynman diagrams showing the (a) $\mathcal O\left(v^0\right)$ and (b) $\mathcal O\left(v^2\right)$ corrections to the quark propagator from loop diagrams mediated by $S_1$ and chiral neutrinos. The couplings $g_L=y_1^{LL}$ and $g_R=y_1^{\overline{RR}}$ [see Eq.~\eqref{eq:Lag}]. The diagrams for $s$- and $b$-quarks are similar. These corrections are independent of the external momentum ($p$) and hence contribute as mass corrections.}
	\label{fig:mass_correction}	 
	\end{center}
	\end{figure}
 \noindent
In our calculation, we assume that left-handed neutrinos are massless while right-handed ones are massive. Also, since we consider Higgs decays to down-type quarks only, we can safely ignore the quark masses ($m_q=0$) and set $m_h^2=(p_1+p_2)^2=2p_1.p_2$ (see Fig.~\ref{fig:lepto}). 
The correction to $y_d$ coming from the diagram shown in Fig.~\ref{fig:leptoa} 
is given by,
\begin{align}
 \tilde y_d^{(a)}=\ -ig_1^2y_\nu\int\frac{d^4\ell}{ (2\pi)^4}\Bigg[&\frac{P_R \slashed{\ell}\ (\slashed{p}_1+\slashed{p}_2-\slashed{\ell}+M_{\nu_R})P_R}{ \ell^2\{ (p_1+p_2-\ell)^2-M_{\nu_R}^2\}}\nonumber\\
 \times\ &\frac1{ \{(p_1-\ell)^2-M_{S_1}^2\}}\Bigg],
\end{align}
where $g^2_1=g_L\,g_R=y_1^{LL}\,y_1^{\overline{RR}}$ and $P_{L/R}$ are the chirality projectors. From here on, we shall suppress the generation index of the leptoquark couplings and simply write $g^2_i$ as $g^2$.
After Feynman parametrization and dimensional regularization, we get
\begin{equation}
\tilde y_d^{(a)}=-\frac{g^2y_\nu}{16\pi^2}\left[\int_0^1 dx\int_0^{1-x}dy\,\left(\frac{x\,m_h^2}{D_1}\right)-\int_0^1 dz\,\ln D_2+\Delta_\epsilon\right],
\end{equation}
where,
\begin{equation}
 D_1(x,y)=xym_h^2+x(x-1)m_h^2+xM_{\nu_R}^2+yM_{S_1}^2,
\end{equation}
and
\begin{equation}
 D_2(z)=zM_{S_1}^2+(1-z)M_{\nu_R}^2.
\end{equation}
The divergent piece at $\mathcal O(v^0)$, $\Delta_\epsilon=\frac{2}{\epsilon}-\gamma+{\rm ln} (4\pi)+ \mathcal O(\epsilon)$ is cancelled by a similar contribution from diagrams with a bubble in an external quark line. The bubble in the quark lines is obtained by replacing the Higgs field in Fig.~\ref{fig:leptoa} with $v$ and amputating the external quark lines, see Fig.~\ref{fig:mass1a}. This extra contribution is given as
\begin{equation}
\left.y_d^{\rm leg}\right|_{\mathcal{O}(v^0)}=\frac{g^2y_\nu}{16\pi^2}\int^1_0 dz\left[\Delta_\epsilon-{\rm ln}\left\{zM_{S_1}^2+ (1-z)M_{\nu_R}^2\right\}\right].
\label{Fig:leg1}
\end{equation}
Putting these two together, we get
\begin{align}
y_d^{(a)}=-\frac{g^2y_\nu}{16\pi^2} \Bigg[&\int_0^1 dx\int_0^{1-x}dy\left (\frac{x\, m_h^2}{D_1}\right)\Bigg].
\end{align}
Now, proceeding along the same lines, we get the correction from the diagram in Fig.~\ref{fig:leptob} as
\begin{align}
y_d^{(b)}=&\ ig^2\lambda y_\nu v^2\int\frac{d^4\ell}{ (2\pi)^4}\Bigg[\frac{1}{ (\ell^2-M_{\nu_R}^2)\{ (\ell-p_1)^2-M_{S_1}^2\}}\nonumber\\
&\qquad\qquad\qquad\quad\times\ \frac1{\{ (\ell+p_2)^2-M_{S_1}^2\}}\Bigg]\nonumber\\
=&\ \frac{g^2\lambda y_\nu v^2}{16\pi^2}\int_0^1 dx\int_0^{1-x}dy\left (\frac{1}{D_0}\right),
\end{align}
where $D_0 (x,y)=M_{\nu_R}^2+ (x+y) (M_{S_1}^2-M_{\nu_R}^2)-xy\, m_h^2$. Similarly, the correction term corresponding to Fig.~\ref{fig:leptoc} can be obtained as,
\begin{align}
 \tilde y_d^{(c)}=&\ \frac{ig^2y_\nu\lambda v^2}{2}\int\frac{d^4\ell}{ (2\pi)^4}\Bigg[\frac{P_R \slashed{\ell}\ (\slashed{p}_1+\slashed{p}_2-\slashed{\ell}+M_{\nu_R})P_R}{ \ell^2\{ (p_1+p_2-\ell)^2-M_{\nu_R}^2\}}\nonumber\\
&\qquad\qquad\qquad\quad\times \ \frac1{ \{(p_1-\ell)^2-M_{S_1}^2\}^2}\Bigg]\nonumber\\
 =&\ -\frac{g^2y_\nu\lambda v^2}{32\pi^2}\Bigg[\int_0^1 dx\int_0^{1-x}dy(1-x-y)\left(\frac{ym_h^2}{D_3^2}\right)\nonumber\\
 &\qquad\qquad\qquad\quad+\ \int_0^1 dx\int_0^{1-x}dy\left(\frac{1}{D_4}\right)\Bigg],
 \label{eq:yctilde}
\end{align}
where $D_3(x,y)=-xy m_h^2+y M_{\nu_R}^2+(1-x-y)M_{S_1}^2$ and $D_4(x)=x M_{\nu_R}^2+(1-x)M_{S_1}^2$. This is finite like $y^{(b)}$. The last term of Eq.~\eqref{eq:yctilde} is actually cancelled by  the $\mathcal O (v^2)$ correction to 
the external quark propagators, as shown in Fig.~\ref{fig:mass1b}. This is similar to the cancellation at $\mathcal O (v^0)$ in $y^{(a)}$: in this case the bubble is obtained by replacing the Higgs field in Fig.~\ref{fig:leptob} with $v$ and amputating the external quark lines. However, one has to be careful with the factors here. After electroweak symmetry breaking, one can expand the Higgs-$S_1$ interaction term  in Eq.~\eqref{eq:lagrangian} as
\begin{align}
\lambda \left(H^\dagger H\right)\left(S_1^\dagger S_1\right)= \frac{\lambda}{2}\left(h^2+2hv+v^2\right)\left(S_1^\dagger S_1\right) + \cdots
\end{align}
The $\lambda v\, h(S_1^\dagger S_1)$ term contributes to $y^{(b)}_q$, but the propagator correction would come from the $\lambda v^2(S_1^\dagger S_1)/2$ term, i.e., with a different prefactor. The $\mathcal O (v^2)$ external leg correction to the Yukawa coupling is proportional to $\lambda v^2(S_1^\dagger S_1)/2$  and can be written as
\begin{align}
 \left.y_d^{\rm leg}\right|_{\mathcal{O}(v^2)}=&\ \frac{g^2y_\nu\lambda v^2}{32\pi^2}\int_0^1 dx\left(\frac{1-x}{x M_{\nu_R}^2+(1-x)M_{S_1}^2}\right).
 \label{Fig:leg2}
\end{align}
Once this is added, we get
\begin{align}
y_d^{(c)}=&\ -\frac{g^2y_\nu\lambda v^2}{32\pi^2}\int_0^1 dx\int_0^{1-x}dy(1-x-y)\left(\frac{ym_h^2}{D_3^2}\right)
 \label{eq:yc}
\end{align}
Therefore, the effective $h d\bar {d}$ coupling can be written as
\begin{align}
y_d^{\rm eff}=&\ y^{\rm SM}_d\,+\, \delta y\nonumber\\
=&\ {y^{\rm SM}_d}
+\frac{g^2y_\nu}{16\pi^2}\Bigg[\int_0^1 dx\int_0^{1-x}dy \, \Bigg\{ \frac{\lambda v^2}{D_0}-\frac{x\,m_h^2}{D_1}\nonumber\\
&\qquad\qquad\quad-(1-x-y)\left(\frac{\lambda v^2 \times ym_h^2}{2D_3^2}\right)\Bigg\}\Bigg],\label{eq:Mtot}
\end{align}
where $y^{\rm SM}_d =m_d/v$ is  the $d$-quark Yukawa coupling in the SM (with $m_d$ being the physical mass) and $\delta y=y_d^{(a)}+y_d^{(b)}+y_d^{(c)}$ is the total loop correction. This results in a finite shift to the SM down-quark Yukawa couplings which cannot be absorbed in a redefinition of the quark masses since the corrections corresponding to the mass terms (Fig. \ref{fig:mass_correction}) 
are already accounted for in $m_d$, the physical mass, through Eqs.~\eqref{Fig:leg1} and~\eqref{Fig:leg2}. 
This is similar to the case in which the SM is augmented with dimension-6 operators~\cite{Bar-Shalom:2018rjs}.  

Equation~\eqref{eq:Mtot} can also be written in terms of the following PV integrals,
\begin{align}
y_d^{\rm eff}\ =&\ \ {y^{\rm SM}_d}+\frac{g^2y_\nu}{16\pi^2}
\Big[B_0 (0,M_{\nu_R}^2,M_{S_1}^2)-B_0 (m_h^2,0,M_{\nu_R}^2)\nonumber\\
&\ - M_{S_1}^2C_0 (0,0,m_h^2,0,M_{S_1}^2,M_{\nu_R}^2)\nonumber\\
&\ - \lambda v^2 C_0 (0,0,m_h^2,M_{S_1}^2,M_{\nu_R}^2,M_{S_1}^2)\nonumber\\
&\ +\frac{\lambda v^2}{2}\Big\{C_0(0,0,m_h^2,0,M_{S_1}^2,M_{\nu_R}^2) \nonumber\\
&\ + M_{S_1}^2 D_0(0,0,m_h^2,0,0,0,M_{S_1}^2,M_{S_1}^2,0,M_{\nu_R}^2)\nonumber\\
&\ - C_0(0,0,0,M_{S_1}^2,M_{S_1}^2,M_{\nu_R}^2)\Big\} \Big],
\label{eq:MtotPV}
\end{align}
where $D_0$, $C_0$ and $B_0$ are the four-point, triangle, and bubble integrals, respectively.
The expressions for the $s$- and $b$-quarks would be exactly the same as the above with $m_d$ and $g^2=g^2_i$ suitably modified.

\begin{table}[t]
\centering
\begin{tabular*}{\columnwidth}{c@{\extracolsep{\fill}}c@{\extracolsep{\fill}}c@{\extracolsep{\fill}}c@{\extracolsep{\fill}}c}
$M_{\nu_R}$ (GeV) & $M_{S_1}$ (GeV) & $y^{(a)}$ & $y^{(b)}$ & $y^{(c)}$\\\hline\hline
\multirow{2}{*}{600} & 1000 & $-0.000046$ & $0.000255$ & $-6.3\times 10^{-7}$\\
 & 1500 & $-0.000031$ & $0.000132$ & $-2.2\times 10^{-7}$\\\hline
\multirow{2}{*}{1100} & 1000 & $-0.000022$ & $0.000180$ & $-2.1\times10^{-7}$\\
 & 1500 & $-0.000016$ & $0.000103$ & $-0.9\times10^{-7}$\\\hline
\end{tabular*}%
\caption{Contributions of the three diagrams shown in Fig.~\ref{fig:lepto} to the Yukawa couplings 
obtained from Eq.~\eqref{eq:Mtot} or~\eqref{eq:MtotPV} for some illustrative choices of the mass of 
the right-handed neutrino $M_{\nu_R}$ and the leptoquark mass $M_{S_1}$ while keeping $g^2y_\nu=1$ and $\lambda=1$.}
\label{tab:loop}
\end{table}
	\begin{figure}[t]
	\begin{center}
\subfloat[\quad\quad\quad (a)]{\includegraphics[width=0.9\columnwidth]{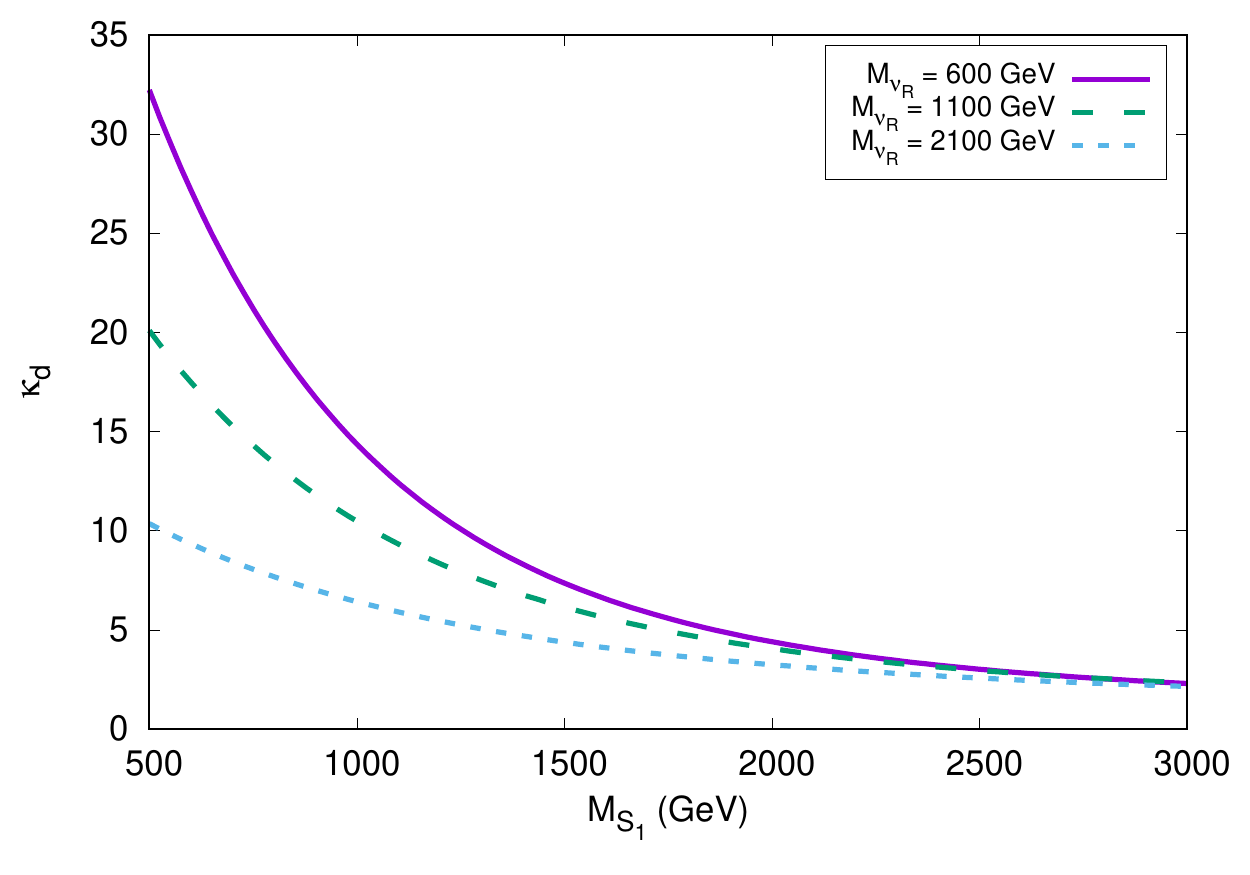}}\\
	\subfloat[\quad\quad\quad (b)]{\includegraphics[width=0.9\columnwidth]{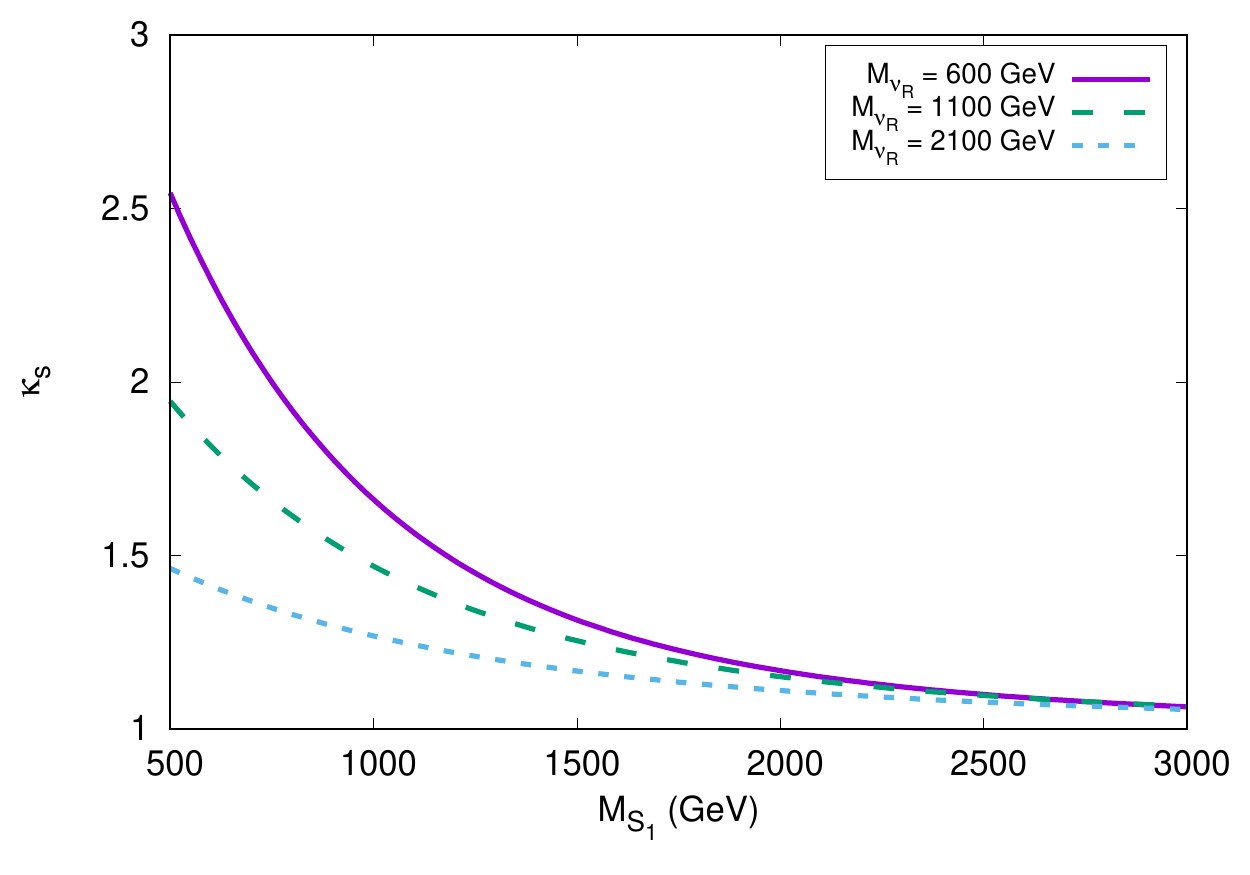}}\\
	\subfloat[\quad\quad\quad (c)]{\includegraphics[width=0.9\columnwidth]{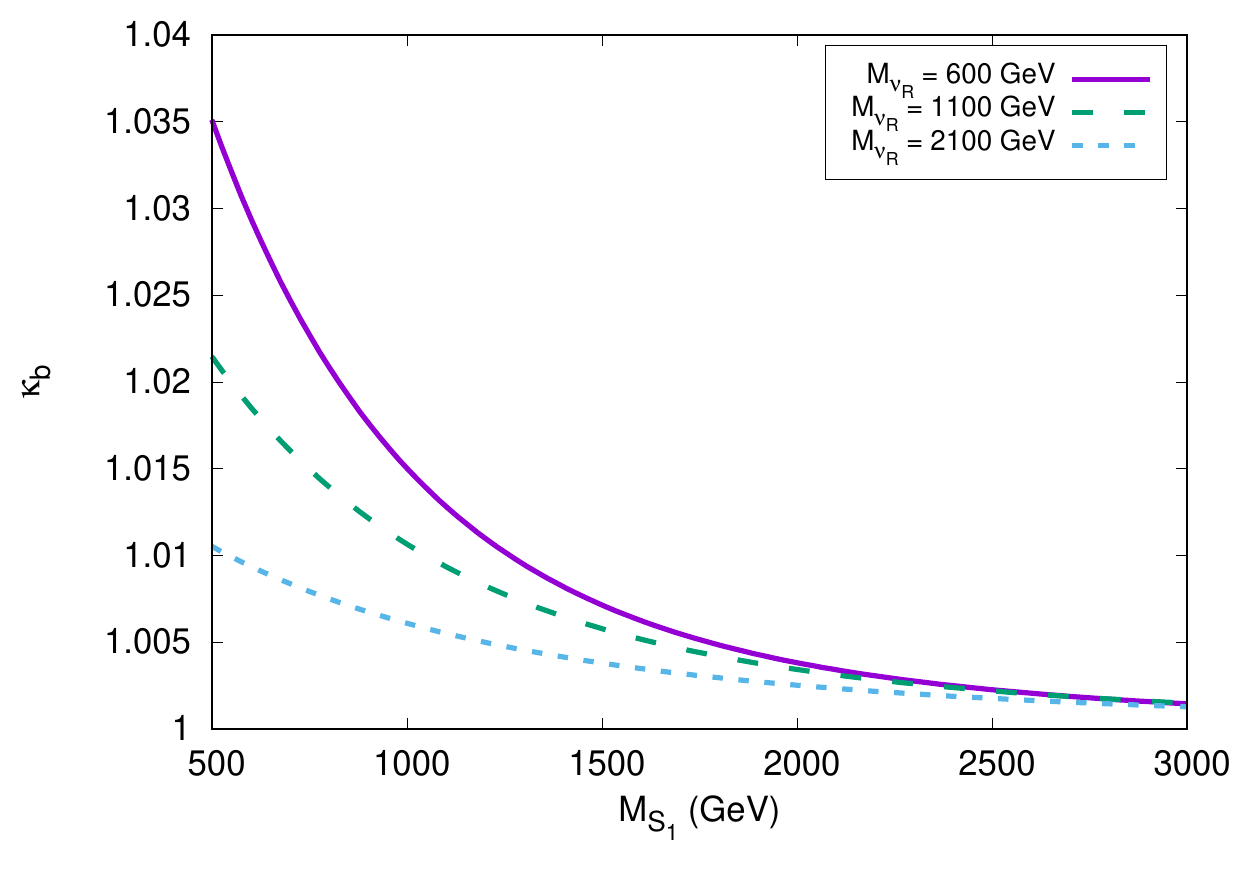}}
\caption{Variation of the coupling modifiers (a) $\kappa_d$, (b) $\kappa_s$, and (c) $\kappa_b$ [defined in Eq.~\eqref{eq:kappa}] with $M_{S_1}$ for different values of $M_{\nu_R}$. Here we set $g^2y_{\nu}=1$ for all three generations and keep $\lambda=1$.}
	\label{fig:Y_tot}	 
	\end{center}
	\end{figure}

\subsection{Relative Couplings}
\noindent
 To get some idea about how the extra contributions from the loops depend on the parameters, we {use the Yukawa coupling modifiers [Eq.~\eqref{eq:kappa}],}
\begin{equation}
\kappa_q=1+\frac{\delta y}{y_q^{\rm SM}}.
\end{equation}
Since we ignore the mass of the quarks, $\delta y$ is 
independent of the flavour of the down-type quark that the Higgs is coupling to as long as $g^2y_\nu$ remains the same. Hence, 
$\delta y/y^{\rm SM}_q$ should go as $1/y^{\rm SM}_q \sim1/m_q$. Using this and Eq.~\eqref{eq:MtotPV},
we see that $\kappa_q$ depends linearly on $1/m_q$, $\lambda$ and the combination $g^2y_\nu$, but, \emph{a 
priori}, its dependence on $M_{S_1}$ or $M_{\nu_R}$ is not so simple. In Table~\ref{tab:loop}, 
we show the contributions of the three loop diagrams [Figs. \ref{fig:leptoa}--\ref{fig:leptoc}] for some illustrative choices of $M_{\nu_R}$ and $M_{S_1}$. With $g^2y_\nu=\lambda=1$, we see that there is some cancellation between these contributions. 
Note that this choice of coupling is not restricted by the
rare decays~\cite {Dorsner:2016wpm,Mandal:2019gff}.

In Fig.~\ref{fig:Y_tot}, we show the variations of $\kappa_d$, $\kappa_s$ and $\kappa_b$ for $500\leq M_{S_1}\leq3000$ GeV for three different choices of $M_{\nu_R}$. As expected, we see the lightest among the three quarks, i.e., the $d$-quark getting the maximum deviation in $\kappa_q$ from unity. The $b$-quark coupling hardly moves from the SM value for the considered parameter range. However, all the deviations are well within the
ranges allowed by Eq.~\eqref{eq:ydlimits}.

\begin{figure}[]
	\subfloat[\quad\quad\quad (a)]{\includegraphics[width=0.9\columnwidth]{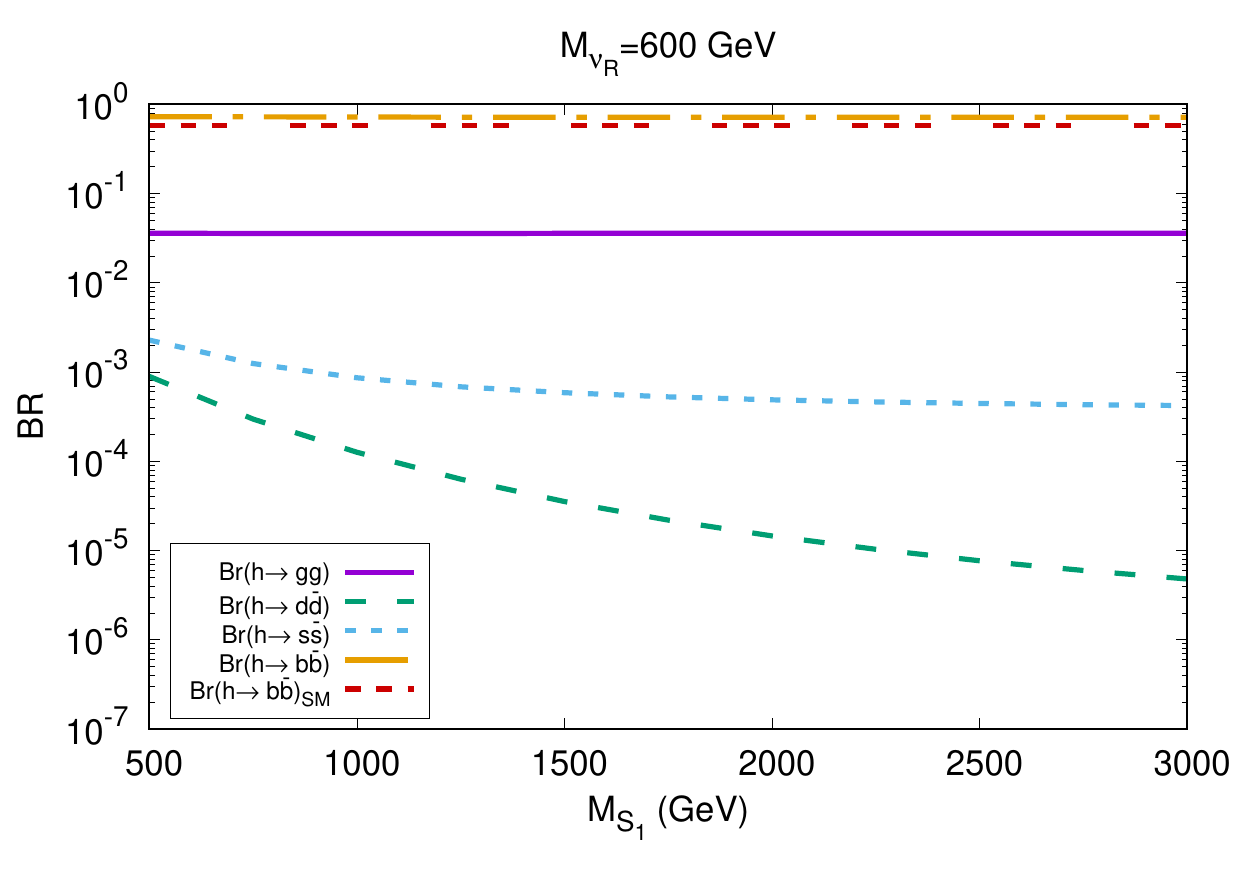}}\\
	\subfloat[\quad\quad\quad (b)]{\includegraphics[width=0.9\columnwidth]{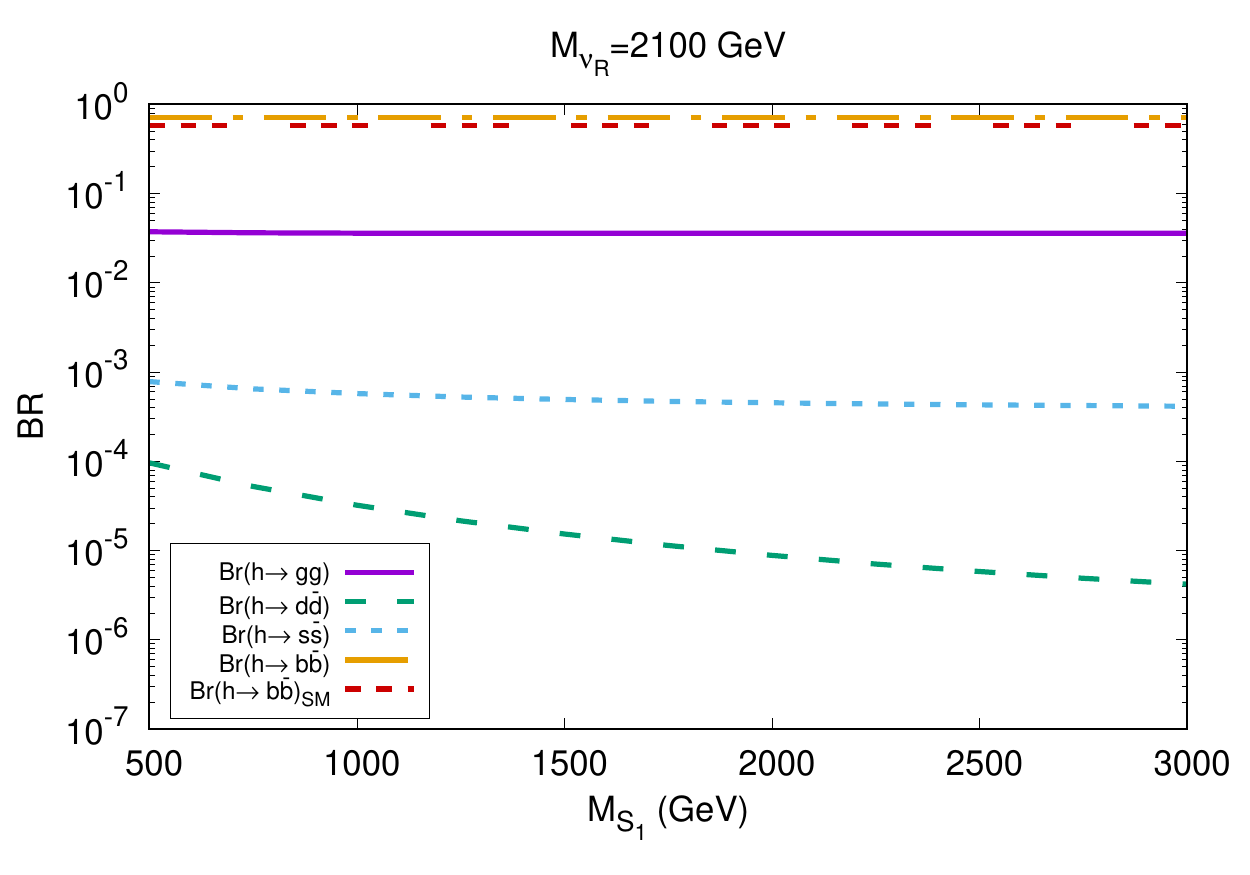}}
	\caption{Variation of BR$(h\rightarrow ii)$ with $M_{S_1}$ for $i=d$, $s$, $b$, and gluon and two values of $M_{\nu_R}$: (a) $600$~GeV and (b) $2100$~GeV. We have set $g^2y_\nu=1$ for all generations and taken $\lambda=1$.}
	\label{fig:Br}
\end{figure} 

\subsection{Decays of $\pmb{h_{125}}$}
\label{subsec:hdecay}
\noindent
As mentioned before, we shall use $h$ and $h_{125}$ interchangeably to denote 
the $125$ GeV SM-like Higgs boson. In the SM, the total decay width of the 125 GeV Higgs boson is computed as $\Gamma^{\rm SM}_h=4.07\times10^{-3}$ GeV, with a relative theoretical uncertainty of $^{+4.0\%}_{-3.9\%}$~\cite{PhysRevD.98.030001}. Now, because of the additional loop contribution, the total decay width would increase in our model. 
We can use Eq.~\eqref{eq:Mtot} or \eqref{eq:MtotPV} to compute the 
partial decay width for the $h\to q\bar q$ decay in the rest frame of 
the Higgs as
\begin{align}
\Gamma_{h\rightarrow q \bar q}&\ =\ N_c\times\frac{|\vec{p_q}|}{32\pi^2m_h^2}\int\left|\mathcal{M}_{tot}\right|^2 d\Omega\nonumber\\
&\ =\ \frac{N_c}{8\pi m_h^2}|y_q^{\rm eff}|^2\left (m_h^2-4m_q^2\right)^{3/2},
\label{eq:hdecay}
\end{align}
where $i\mathcal{M}_{tot}=y_q^{\rm eff} {q} \bar{q}$ is the invariant amplitude and $N_c=3$ accounts for the colours of the quark. Similarly, the $h\to gg$ partial width would also get a positive boost
in the presence of $S_1$~\cite{Dorsner:2016wpm}. The relevant diagrams can be seen in Figs.~\ref{fig:glu_a} and~\ref{fig:glu_b}. In our model, the $h\to gg$ partial width can be expressed as~\cite{Djouadi:2005gj,Dorsner:2016wpm},
\begin{align}
 \Gamma_{h\rightarrow gg}&=\frac{G_{\rm F}\alpha_s^2m_h^3}{64\sqrt{2}\pi^3}\left|\mathcal{A}_{1/2}(x_t)+\frac{\lambda v^2}{2M_{S_1}^2}\mathcal{A}_0(x_{S_1})\right|^2
 \label{eq:gammaSM_h}
\end{align}
where $x_t=m_h^2/4m_t^2$ and $x_{S_1}=m_h^2/4M_{S_1}^2$. The 
relevant one-loop functions are given by
\begin{align}
 \mathcal{A}_{1/2}(x)&=\frac{2[x+(x-1)f(x)]}{x^2}\,,\\
 \mathcal{A}_0(x)&=-\frac{[x-f(x)]}{x^2}\,,\\
 f(x)&=\left\{\begin{array}{cc}
 {\rm arcsin}^2(\sqrt{x}), & x\leq 1\\
 -\frac{1}{4}\left[{\rm ln}\left(\frac{1+\sqrt{1-x^{-1}}}{1-\sqrt{1-x^{-1}}}\right)-i\pi\right]^2, & x>1
 \end{array}\right\}.
 \label{eq:glu_func}
\end{align}
Now, Eqs.~\eqref{eq:hdecay} and~\eqref{eq:gammaSM_h} can be used to obtain the total width in our model, 
\begin{align}
\Gamma_h=&\left (\Gamma_h^{\rm SM}-\Gamma^{\rm SM}_{h\rightarrow gg}-\sum_{q=d,s,b}\Gamma^{\rm SM\, (tree)}_{h\rightarrow q \bar q}\right)+\Gamma_{h\rightarrow gg}\nonumber\\
&+\sum_{q=d,s,b}\Gamma_{h\rightarrow q \bar q}.
\label{eq:decay}
\end{align} 
Ideally, we should also include corrections to partial widths of other decay modes, like $h\rightarrow \gamma\gamma$ or other three body decays in the above expression. However, since their contributions to the total width are relatively small, we ignore them.

From Eqs.~\eqref{eq:hdecay} and~\eqref{eq:decay}, we compute the new branching ratios (BRs) of the $h\to q \bar q$ modes in our model as
\begin{align}
 {\rm BR}(h\to q \bar q) = \frac{\Gamma_{h\to q \bar q}}{\Gamma_h}. 
\end{align}
 In Fig.~\ref{fig:Br} we show BR$(h\rightarrow q\bar{q})$ for different 
 quarks for $g^2y_\nu=1$ (for all generations) and $\lambda=1$. 
Equation~\eqref{eq:hdecay} indicates 
BR$(h\to q \bar q)\sim |y_q^{\rm SM}+\delta y|^2$, i.e., it increases with $y_q^{\rm SM}$ (remember, for $g^2y_\nu=1$, $\delta y$ is the same for all the quarks). Hence, we expect BR$(h\to b \bar b)>$ BR$(h\to s \bar s)>$ BR$(h\to d \bar d)$, as $y_q^{\rm SM}$ increases with the mass of the quark. This
can be seen in Fig.~\ref{fig:Br}. However, even for order $1$ $y_\nu$ couplings and TeV 
 scale $S_1$ and $\nu_R$, the relative shift in branching ratio of the $h\to b \bar b$ decay to that of SM 
 is not large (as expected from Fig.~\ref{fig:Y_tot}). For the lighter quarks, the branching ratios become much larger than their SM values, even though they remain small compared to other decay modes like $h\rightarrow b\bar{b}$. The branching fraction
 $h\rightarrow gg$ is almost unaffected with the variation in $S_1$, as the SM contribution always dominates.

\subsection{Production of $\pmb{h_{125}}$}
\noindent
For a quantitative understanding of the quark-gluon fusion production of $h_{125}$, we normalise the fusion cross section with respect to its SM value. We define the ``normalized production" factor $\mu_{\rm F}$ as
\begin{align}
\mu_{\rm F}\equiv\mu_{\rm F}^{gg+q\bar q}&=\frac{\sigma (gg\rightarrow h)+\sum_{q=d,s,b}\sigma (q \bar q\rightarrow h)}{\sigma ( gg\rightarrow h)_{\rm SM}}.
\end{align}
It is a function of the BSM parameters and
measures the relative enhancement of production cross section in the fusion channel.
The subscript ``F'' stands for the fusion channel. In the denominator, we ignore $\sigma(b\bar{b}\to h)_{\rm SM}$, as it is much smaller than $\sigma(gg\to h)_{\rm SM}$ because of the small $b$-quark parton distribution function (PDF) in the initial states. 
\begin{figure}[t]
	\subfloat[\quad\quad\quad(a)]{\includegraphics[width=0.9\columnwidth]{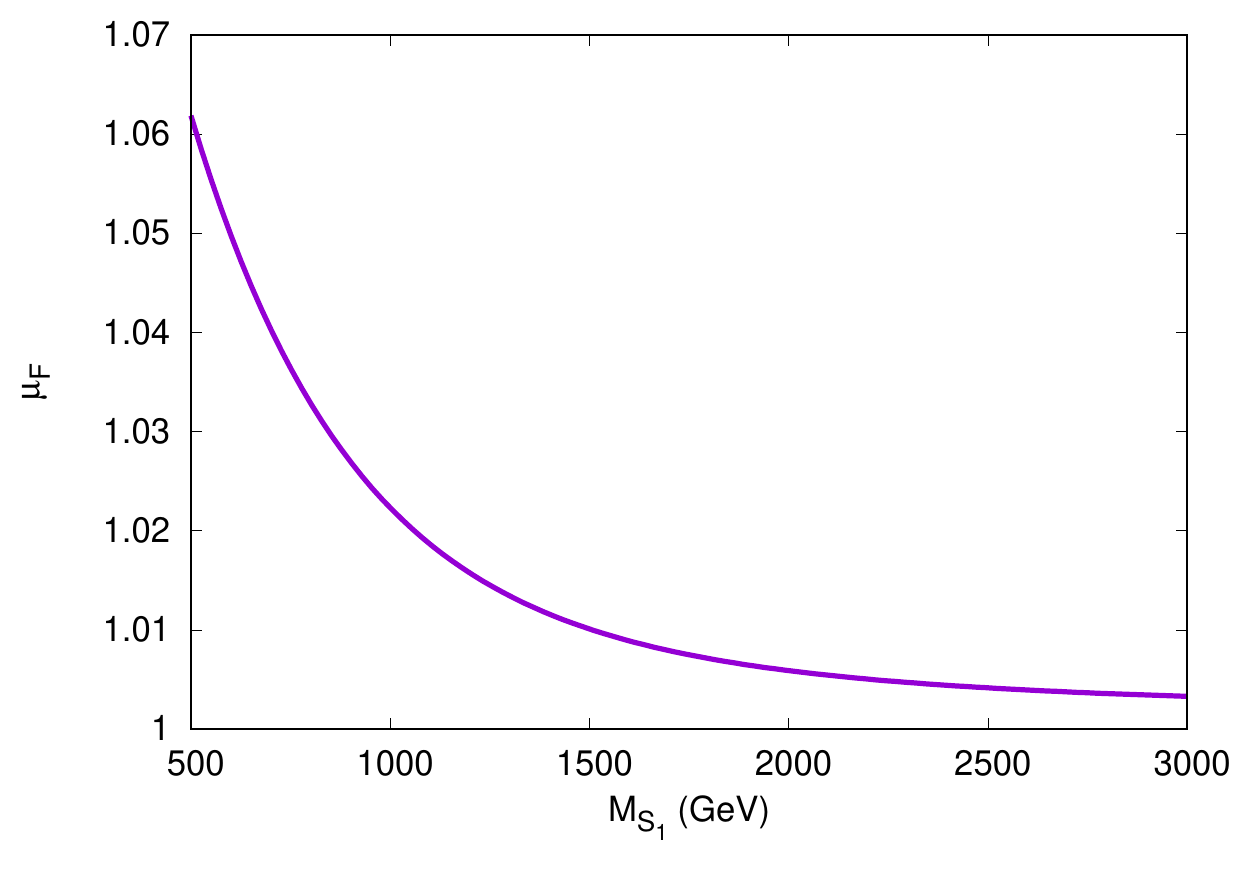}\label{fig:mua}}\\
	\subfloat[\quad\quad\quad(b)]{\includegraphics[width=0.9\columnwidth]{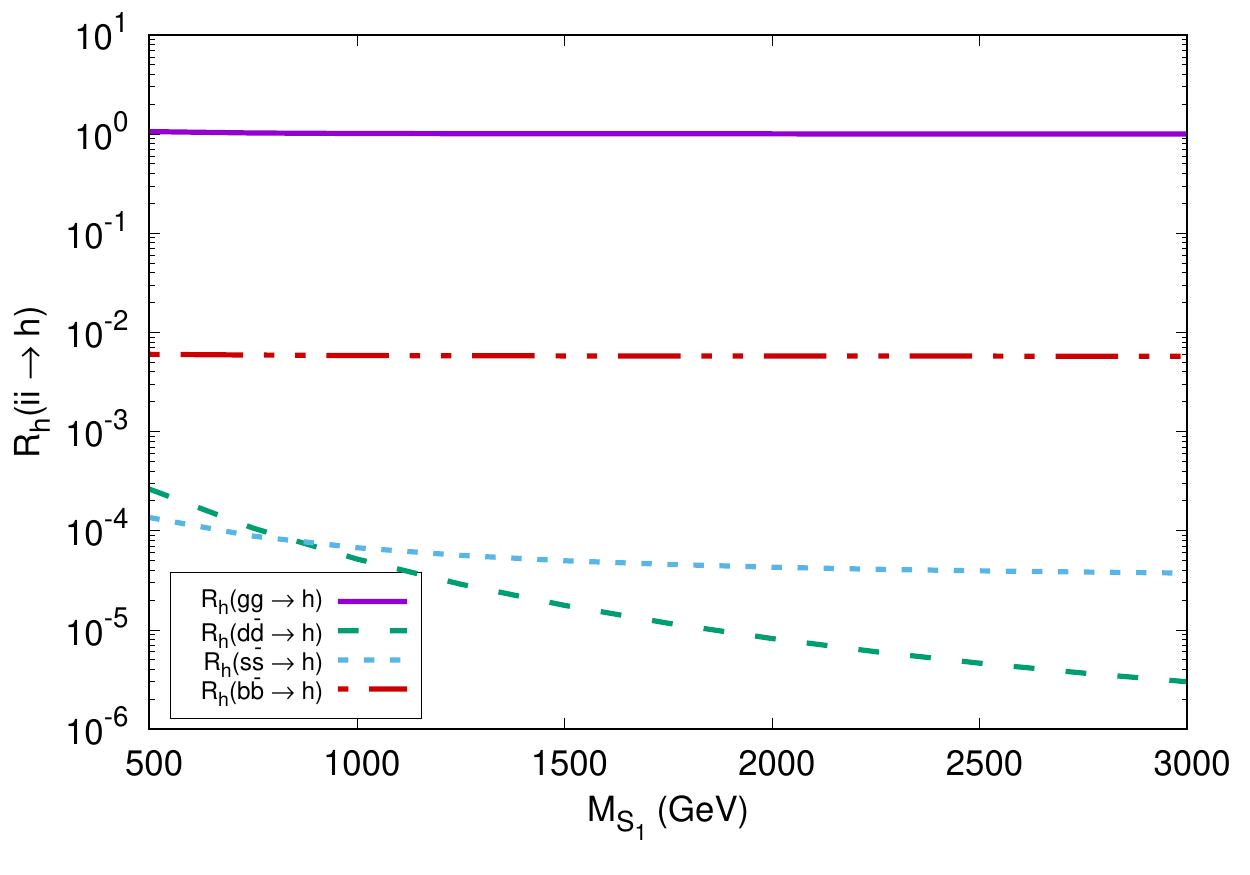}\label{fig:mub}}\\
	\subfloat[\quad\quad\quad(c)]{\includegraphics[width=0.9\columnwidth]{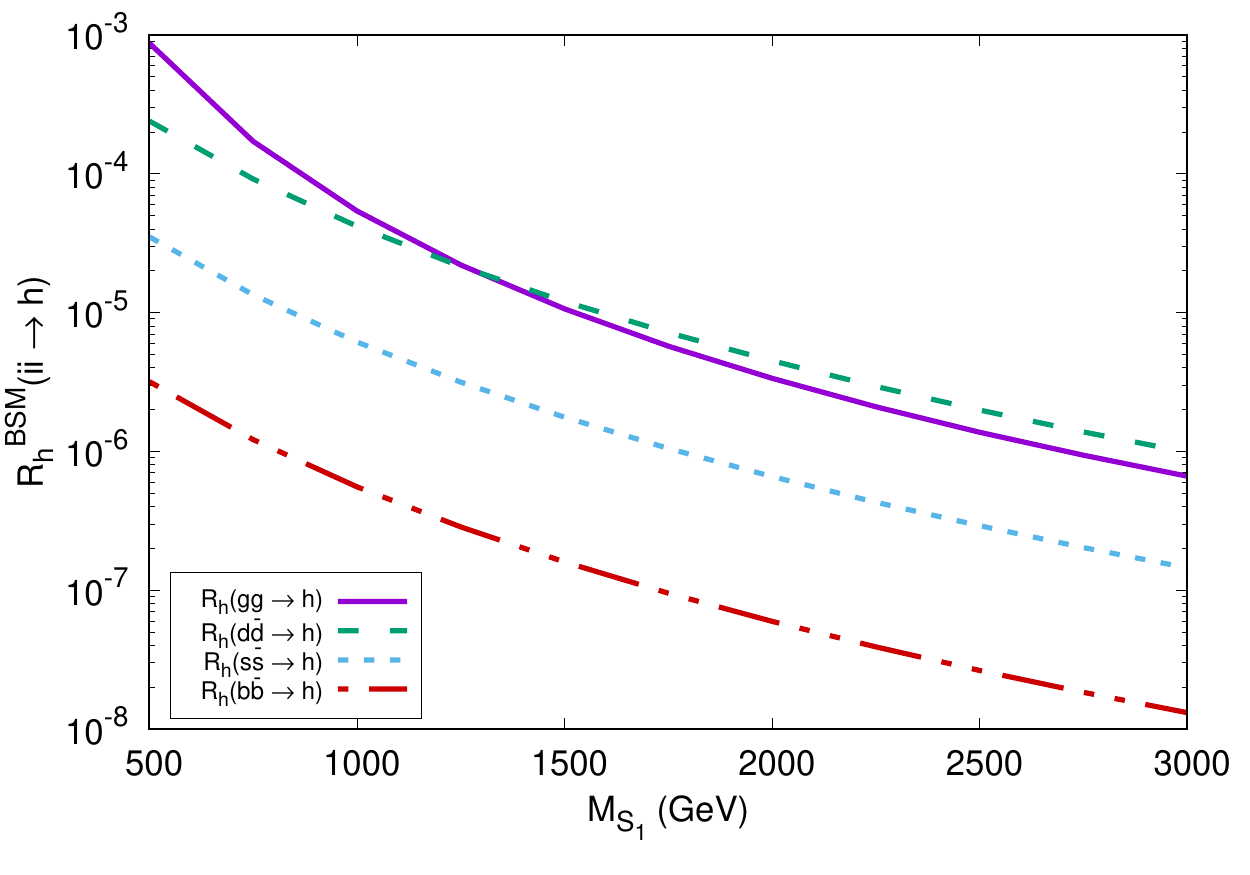}\label{fig:muc}}
\caption{ (a) The normalized production cross section of $h_{125}$ as a function of $M_{S_1}$ for $M_{\nu_R}=1.1$ TeV. Here 
also, we take $g^2y_\nu=1$ for all the generations and $\lambda=1$. (b) Relative production factor~(${\rm R_h}$) (defined in the text) for the SM+LQ scenario as a function of $M_{S_1}$ for $M_{\nu_R}=1.1$ TeV. (c) Relative production factor~$({\rm R_h^{BSM}})$ as a function of $M_{S_1}$ for $M_{\nu_R}=1.1$ TeV when the $hq\bar{q}$~($hgg$) coupling at leading order in the SM is assumed to be zero.}
	\label{fig:mu}
\end{figure}

In our model, the leading order gluon fusion cross section at parton level 
can be expressed as~\cite{Dorsner:2016wpm,Gunion:1986nh,Djouadi:2005gj,Chang:2012ta}
\begin{align}
 \hat{\sigma}(gg\to h)=\frac{\pi^2 m_h}{8\hat s}\Gamma_{h\to gg}\delta(\hat{s}-m_h^2),\label{eq:gghprod}
 \end{align}
 where $\Gamma_{h\to gg}$ is given in Eq.~\eqref{eq:gammaSM_h}. Similarly,
the quark fusion cross section at parton level can be expressed in terms of $\Gamma_{h\to q \bar q}$ from
Eq.~\eqref{eq:hdecay} as~\cite{PhysRevD.98.030001}
\begin{equation}
\hat{\sigma} (q\bar{q}\to h)=\frac{4\pi^2m_h}{9\hat s}\Gamma_{h\to q\bar q}\delta(\hat{s}-m_h^2) \label{eq:hprod}.
\end{equation} 
Naively, one would expect $\hat \sigma(q \bar q\to h)$ for the heavier quarks to be larger than the lighter ones, as $\Gamma_{h\to q\bar q}$ is proportional to the square of 
$y^{\rm eff}_q$ (which increases linearly with $m_q$).
However, there is a trade-off between $m_q$ and the PDFs, as the heavier quarks PDFs are suppressed compared to their lighter counterparts.
We compute $\sigma(q\bar{q}\to h)$ at the 14 TeV LHC using the NNPDF2.3QED LO~\cite{Ball:2013hta} PDF. Similarly, we use the next-to-next-to-leading-order plus next-to-next-to-leading-logarithmic (NNLO+NNLL) QCD 
prediction for the 14 TeV LHC which leads to $\sigma(gg\rightarrow h)_{\rm SM}\simeq49.47$ pb \cite{Higgs14}. We use these results to compute 
$\mu_{\rm F}$. 
We show $\mu_{\rm F}$ as a function of $M_{S_1}$ in Fig.~\ref{fig:mua} while assuming that $g^2y_\nu=1$ for
all the generations and $\lambda=1$. For this plot, we set $M_{\nu_R}=1$ TeV. However, since the gluon fusion cross section is much larger than the quark fusion ones, $\mu_{\rm F}$ is largely insensitive to $M_{\nu_R}$.

To get an idea of the contributions of the different modes to $\mu_{\rm F}$, we define the following two ratios
\begin{align}
 R_h(ii\to h)~=&~\frac{\sigma(ii\to h)}{\sigma(gg\to h)_{\rm SM}}\quad{\rm (full~model)},
 \label{eq:R_h}\\
 R_h^{\rm BSM}(ii\to h)~=&~\frac{\sigma(ii\to h)_{\rm BSM}}{\sigma(gg\to h)_{\rm SM}}\quad{\rm (BSM~only)}\label{eq:R_h_LQ}.
 \end{align}
The difference between these two ratios lies in the interference between the SM and BSM contributions.
We show these ratios in Figs.~\ref{fig:mub} and~\ref{fig:muc}. We find that,
even after the PDF suppression, $R_h(b \bar b\to h)>R_h(s \bar s\to h)>R_h(d \bar d\to h)$. 
On the other hand,
if we take $R_h^{\rm BSM}$, the hierarchy is reversed. This can be understood from the fact that
the loop contribution $\delta y$ is equal for all three of the quarks and hence the PDF suppression makes $R_h^{\rm BSM}(b \bar{b}\to
h)<R_h^{\rm BSM}(s \bar{s}\to h)<R_h^{\rm BSM}(d \bar {d}\to h)$.
Of course, because of the large gluon PDF, $\sigma\left(gg\to h\right)$ is larger than any quark fusion cross section.

\section{Limits on Parameters}\label{sec:limit}
\noindent 
Any increase in either the productions or the decays of $h_{125}$ would be constrained by the existing measurements~\cite{Aad:2019mbh} (also see~\cite{Cepeda:2019klc} for future projections). However we see from Figs.~\ref{fig:Br} and~\ref{fig:mu} that the parameters we consider, i.e., $g_i^2=y^{LL}_iy^{\overline{RR}}_i =1$, $y^{RR}_i =0$, $\lambda=1$, $y_\nu=1$, and TeV scale $M_{S_{1}},\, M_{\nu_{R}}$ for all three generations are quite
consistent with the present and future $h_{125}$ limits.

Concerning the bounds on $S_1$, we see that in our parameter region of interest, 
LQ $S_1$ can decay to all the SM fermions. According to Eqs.~\eqref{eq:lagrangian} and~\eqref{eq:Lag}, a heavy 
$S_1$ would have six decay modes for $M_{S_1}\leq M_{\nu_R}$,
\begin{equation}
S_1\to \left\{ue , c\mu , t\tau , d\nu , s\nu , b\nu \right\},
\end{equation}
with roughly equal BR ($\sim 1/6$) in each mode (if we ignore the differences among the masses of the decay products in different modes). 
The LHC has put exclusion bounds on 
scalar leptoquarks in the light-leptons+jets ($\ell \ell jj/\ell\nu jj$)
\cite{Aaboud:2019jcc,Sirunyan:2018btu,Sirunyan:2018ryt} and $bb\nu\nu/tt\tau\tau$ 
\cite{Sirunyan:2018kzh,Sirunyan:2018ruf,Aaboud:2019bye} channels (see also~\cite{Takahashi:2019zsl,Wong:2019sxu}). The strongest exclusion
limit ($\sim 1.5$ TeV) comes from the $\ell \ell jj$ channel for $100$\% BR in the $S_{1}\to \ell j$ decay. These searches are for pair production of scalar leptoquarks, where the observable signal cross sections are proportional to the square of the BR involved. Hence, in our case, the limit on S$_{1}$ would get much weaker. A conservative estimation indicates that the limit goes below a TeV when the BR decreases to about $1/6$. Also, pair productions of leptoquarks are QCD driven and thus cannot be used to put limits on the fermion couplings.
The CMS Collaboration has performed a search with the $8$ TeV data for single production of scalar leptoquarks that excludes up to 
$1.75$ TeV 
for order $1$ coupling to the first generation \cite{Khachatryan:2015qda}. However, even that limit comes down below $1$ TeV once 
we account for the reduction in the BR. However, a recast of CMS $8$ TeV data for the first 
generation ($eejj/e\nu jj$) indicates that for order $1$ $g_{(L/R)}$, $M_{S_1} 
\gtrsim 1.1$ TeV~\cite{Mandal:2015vfa}.\footnote{Recasting limits from the 
single production searches is trickier than the pair production case because 
here the production processes also depend on the unknown couplings. Even though the parton-level cross section scales easily with 
these couplings, one cannot account for the PDF variation for different quarks in such a simple manner. Since we are interested in a 
conservative limit, we have ignored the PDF variation to obtain this number.} To be on the conservative side, we may use 
$M_{S_1}\gtrsim 1.5$ TeV as a mass limit for $S_1$ with $g^2y_\nu=1$ for all generations.

If, however, $M_{S_1}>M_{\nu_R}$, the LQ can decay to three more final states with right-handed neutrinos. Thus, we would expect a further reduction of the limits on $S_1$~\cite{Das:2017kkm}.
Moreover, specifically for first generation fermions, the choices of $g_L$ and $g_R$ are restricted further. The atomic parity violation 
measurements in ${\rm Cs}^{133}$~\cite{LANGACKER1991277} put a strong constraint on them. Typically, all existing constraints may be satisfied 
easily for $M_{S_1}\gtrsim 2$~TeV and $g^2\approx 1$ with $g_L = g_R$.

\begin{figure}[b]
	\includegraphics[scale=0.5]{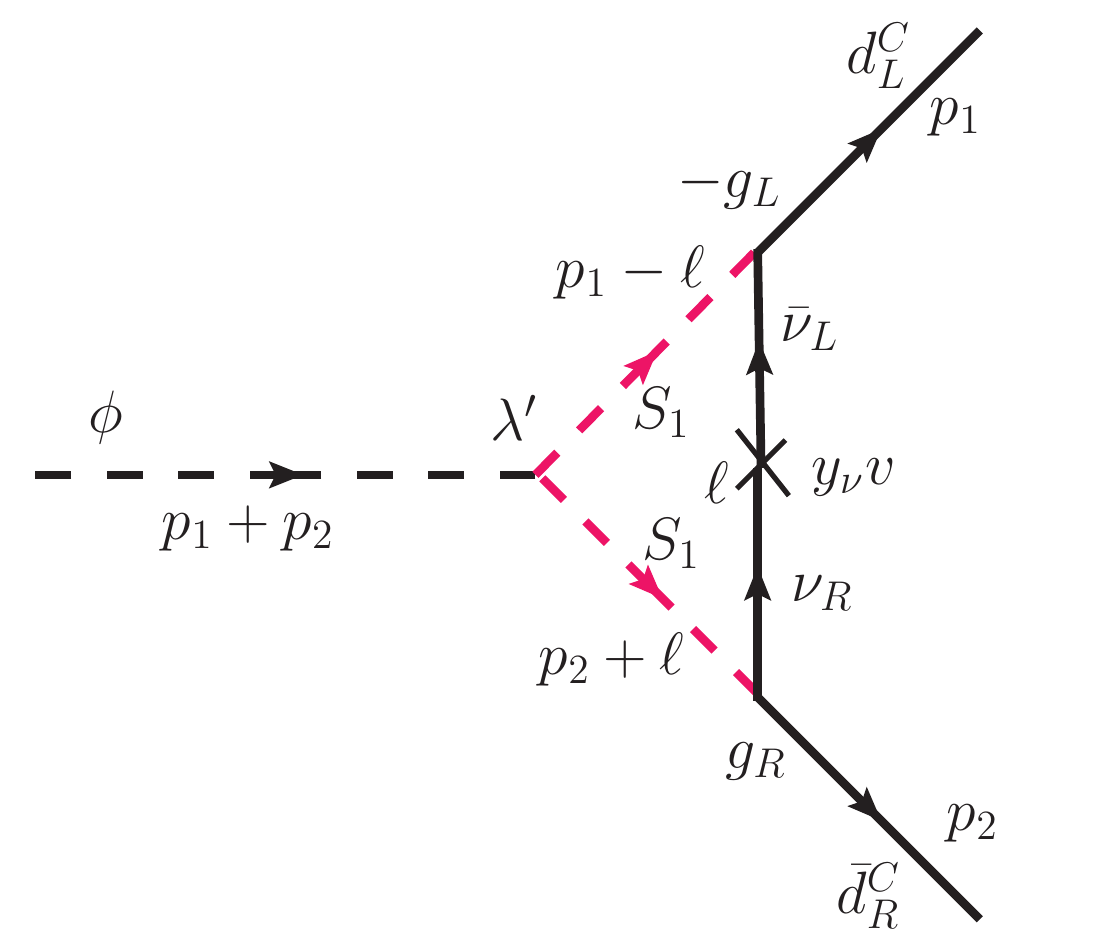}
	\caption{The singlet scalar, $\phi$ decaying to down-type quarks.}
	\label{fig:lep_Phi}
\end{figure}


\section{The Singlet Higgs $\phi$}\label{subsec:analytic_phi}
\noindent Unlike the case of $h_{125}$, the parameters of the singlet scalar defined in Eq.~\eqref{eq:lagrangian} are largely unconstrained.
Generally, to probe a heavy BSM scalar, its decays to fermion pairs like $\tau\tau$ or the massive gauge bosons are assumed to be promising. But, for a singlet scalar, these decay modes lose importance. Also, most of the BSM singlet scalar searches rely on the mixing among the singlet state with the doublet one(s), either $h_{125}$ or other BSM heavy Higgs states.
In our model, by contrast, $\phi$'s can be produced from and decay to a pair of gluons or quarks via the loop of $S_1$ 
and neutrinos without relying, in general, on the mixing of $\phi$ with the doublet Higgs. 
Hence, its phenomenology at the hadron collider would be different than what is generally considered in the literature.
\begin{figure}[]
	\includegraphics[width=0.9\columnwidth]{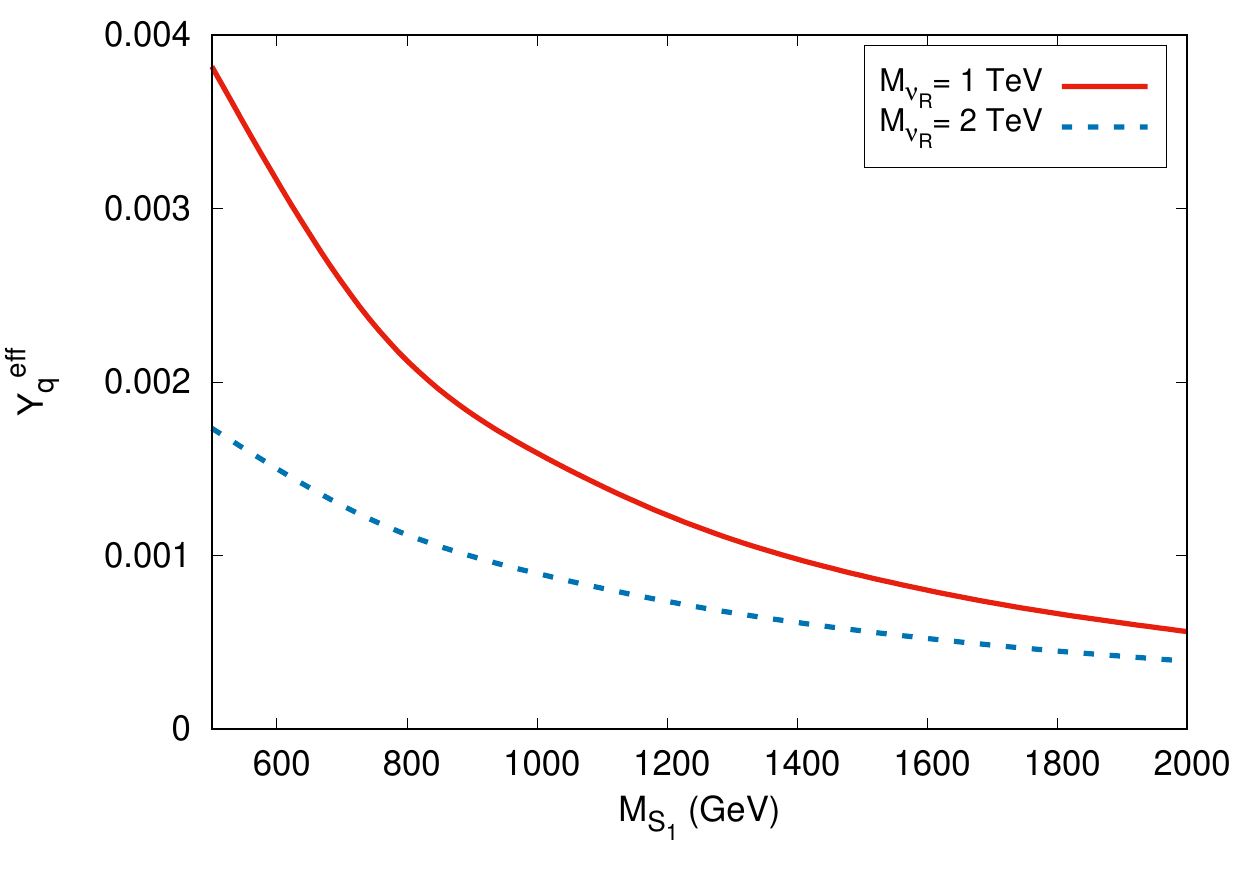}
	\caption{Variation of $\phi q\bar{q}$ coupling as a function of $M_{S_1}$ 
	for $M_{\nu_R}=1$ TeV and 2 TeV, 
	$g^2y_\nu=1$, $\lambda^\prime=2$ TeV and $M_\phi=500$ GeV. }
	\label{fig:sing_Yq}
\end{figure}

\subsection{Effective Coupling}\noindent 
We first calculate the effective couplings of $\phi$ to the light quarks, 
as we did for $h_{125}$. The $\phi q\bar{q}$ effective coupling $Y_q^{\rm eff}$ (where $q$ is any down-type quark) would receive contribution from diagrams like the one 
shown in Fig.~\ref{fig:lep_Phi}, which is similar to the one shown in Fig.~\ref{fig:leptob}. Because of the singlet nature of $\phi$, the tree-level $\phi \bar{\nu}_L \nu_{R}$ coupling does not exist, so in this case, there is no diagram 
like the one shown in Fig.~\ref{fig:leptoa}. Proceeding as before, we get
\begin{align}
Y_{q}^{\rm eff}=\frac{g^2\lambda^\prime y_\nu v}{16\pi^2}\int_0^1 dx\int_0^{1-x}dy\left (\frac{1}{D_\phi}\right),\nonumber
\label{eq:lep_Phi}
\end{align}
where
\begin{equation}
 D_\phi (x,y)=M_{\nu_R}^2+ (x+y) (M_{S_1}^2-M_{\nu_R}^2)-xy\,M_\phi^2.
\end{equation}
Written in terms of PV integrals, this becomes
\begin{align}
Y_{q}^{\rm eff}=-\frac{g^2\lambda^\prime y_\nu v}{16\pi^2} C_0 (0,0,M_\phi^2,M_{S_1}^2,M_{\nu_R}^2,M_{S_1}^2).
\label{Yq_PV}
\end{align}

We present our results in Fig.~\ref{fig:sing_Yq}, which shows the variation of $Y_q^{\rm eff}$ as a function of $M_{S_1}$ for two values of $M_{\nu_R}$ and $M_\phi=500$ GeV. 
Here, $\lambda^\prime$ is a dimensionful parameter [see Eq.~\eqref{eq:lagrangian}] that can be taken to be of the order of the largest mass in the model spectrum. The coupling $Y_q^{\rm eff}$ decreases as $M_{S_1}$ increases.
Since $\phi$ has only loop-level interaction with the SM quarks, the effective coupling is the same for all three generations of down-type quarks for the same value of $g^2y_\nu$. 

\begin{figure}
	\subfloat[\quad\quad\quad (a)]{\includegraphics[width=0.9\columnwidth]{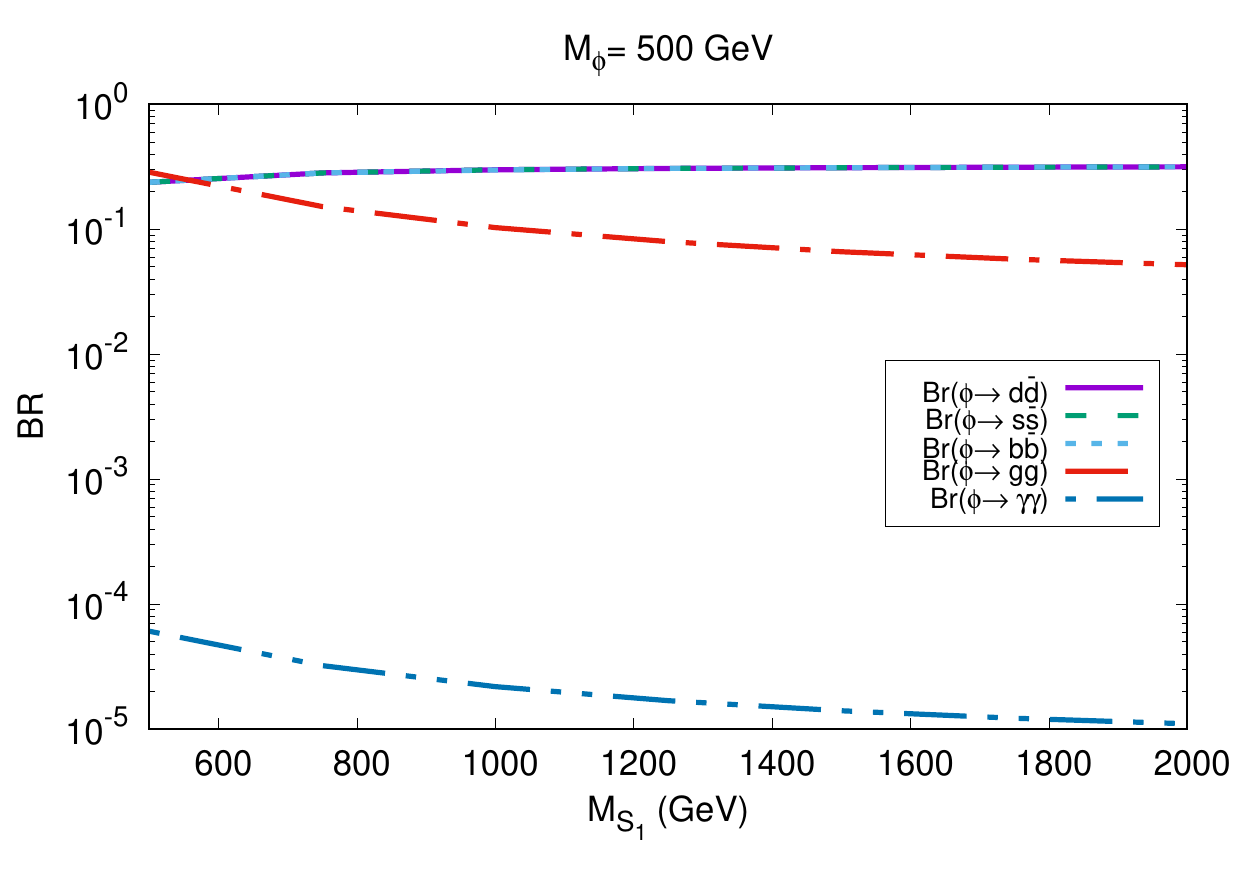}}\\
	\subfloat[\quad\quad\quad (b)]{\includegraphics[width=0.9\columnwidth]{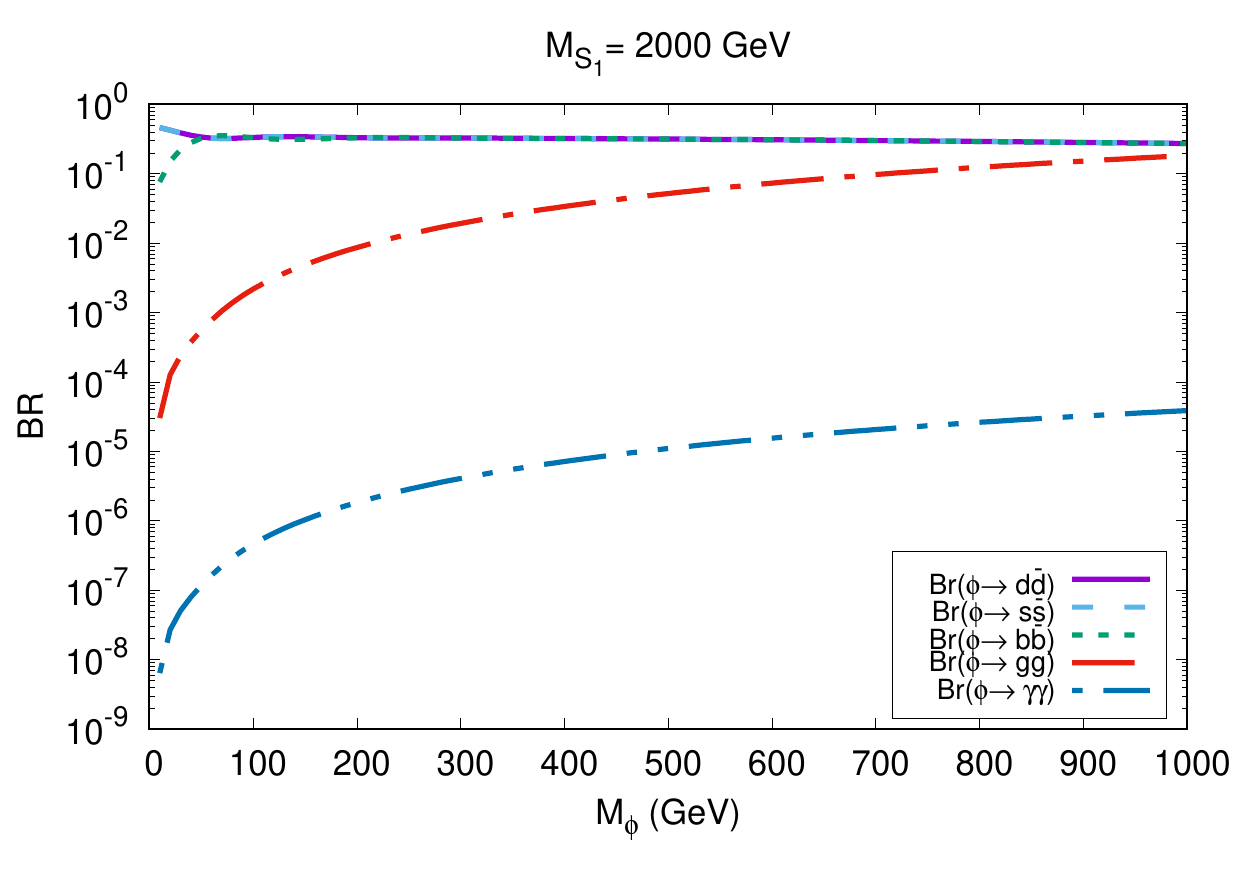}}
	\caption{Variation of Br $(\phi\rightarrow q\bar{q})$, Br $(\phi\rightarrow gg)$ and Br $(\phi\rightarrow \gamma\gamma)$ as a function of (a) $M_{S_1}$ and (b) $M_\phi$ for $g^2y_\nu=1$ and $M_{\nu_R}= 1$ TeV. The ratios are independent of $\lambda^\prime$. We set $\lambda^\prime=2$ TeV to compute the partial decay widths of $\phi$.}
	\label{fig:sing_Br}
\end{figure}

\begin{figure}
 \subfloat[\quad\quad\quad (a)]{\includegraphics[width=0.9\columnwidth]{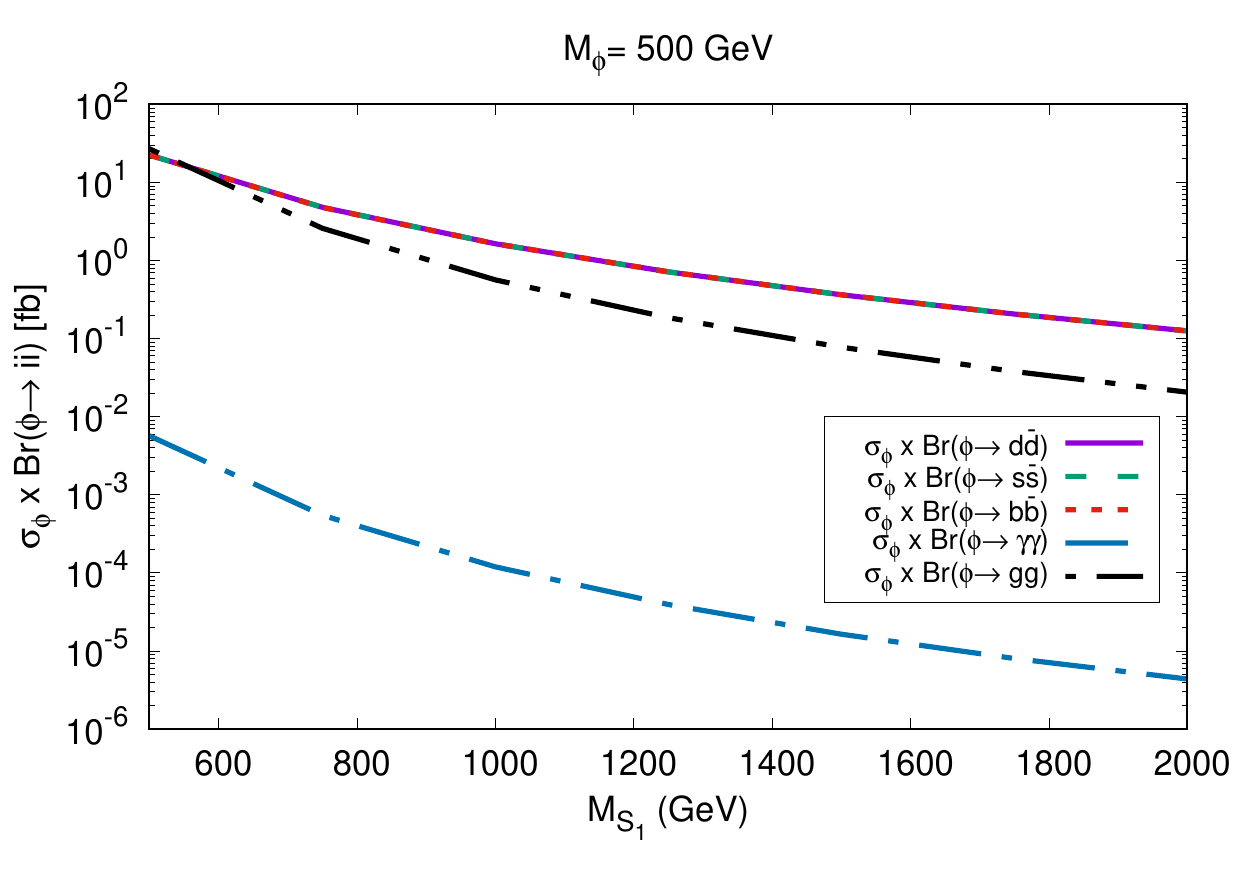}\label{fig:sing_BrCsa}}\\
	\subfloat[\quad\quad\quad (b)]{\includegraphics[width=0.9\columnwidth]{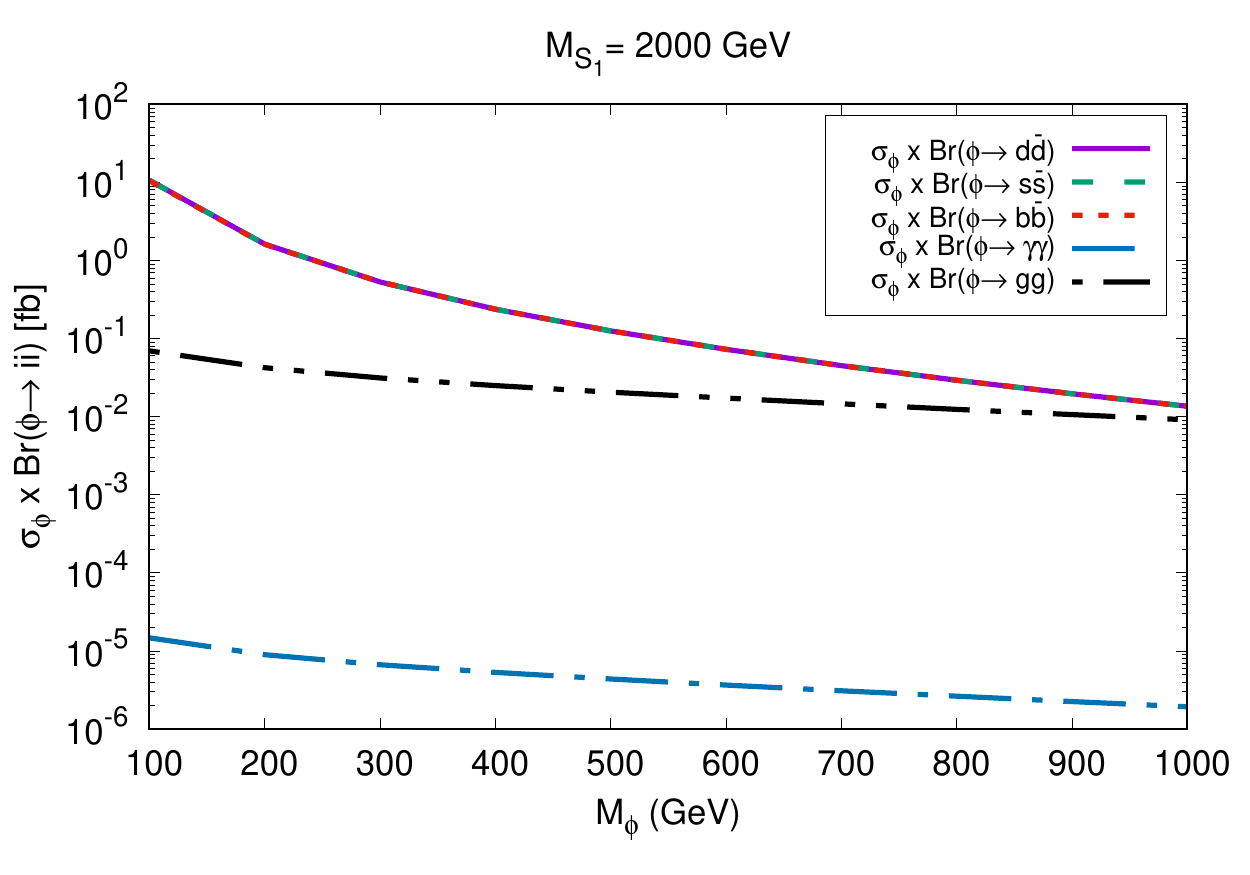}\label{fig:sing_BrCsb}}
	\caption{Variation of the production cross section of $\phi$ times the branching ratios as functions of (a) $M_{S_1}$ and (b) $M_\phi$ for $g^2y_\nu=1$, $\lambda^\prime =2$ TeV and $M_{\nu_R}= 1$ TeV at the 14 TeV LHC.}
	\label{fig:sing_BrCs}
\end{figure}

\begin{figure}
 \subfloat[\quad\quad\quad (a)]{\includegraphics[width=0.9\columnwidth]{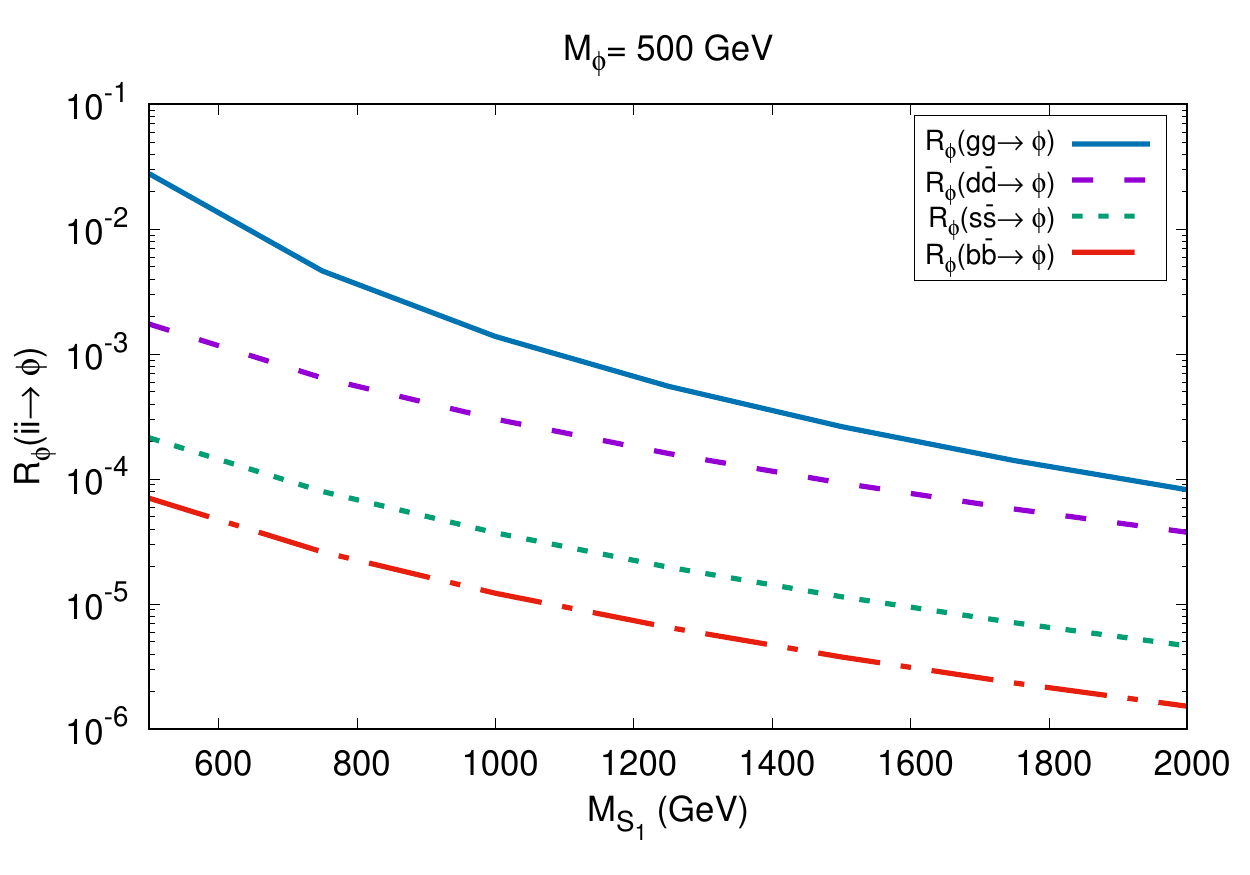}\label{fig:sing_Redgluona}}\\
	\subfloat[\quad\quad\quad (b)]{\includegraphics[width=0.9\columnwidth]{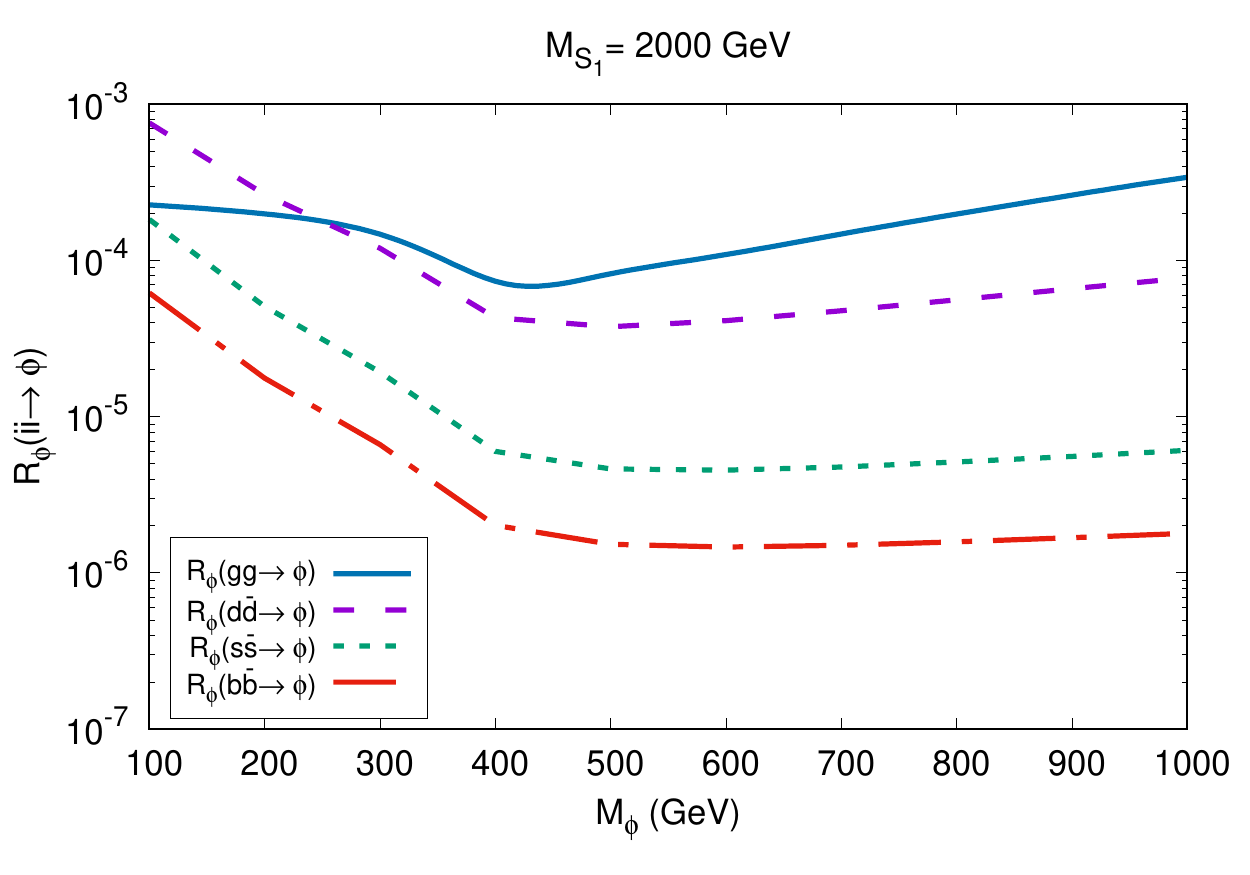}\label{fig:sing_Redgluonb}}
	\caption{Variation of $R_\phi (ii\to\phi)$ [Eq.~\eqref{eq:redsigma}] with (a) $M_{S_1}$ and (b) $M_\phi$ for $g^2y_\nu=1$, $\lambda^\prime=2$ TeV and $M_{\nu_R}=1$ TeV.}
	\label{fig:sing_Redgluon}
\end{figure}

\subsection{Branching Ratios and Cross Sections}
\noindent The expressions for the partial decay widths and production cross section of $\phi$ are essentially 
identical to the ones for $h_{125}$ if we replace $y^{\rm eff}_q\to Y_q^{\rm eff}$ and {$m_h\to M_\phi$}.
Thus, the expressions for the partial decay widths would look like
\begin{align}
\Gamma_{\phi\rightarrow q \bar q}&\ =\ \frac{3|Y_q^{\rm eff}|^2}{8\pi M_\phi^2}\left (M_\phi^2-4m_q^2\right)^{3/2}\approx \frac{3}{8\pi}|Y_q^{\rm eff}|^2M_\phi,\label{eq:singlet-phi-gg}\\
\Gamma_{\phi\rightarrow gg}&\ =\ \frac{G_{\rm F}\alpha_S^2M_\phi^3}{64\sqrt{2}\pi^3}\left|\frac{\lambda^\prime v }{2M_{S_1}^2}\mathcal{A}_0\left(\frac{M_\phi^2}{4M_{S_1}^2}\right)\right|^2, \label{singlet-phi-gg}\\
\Gamma_{\phi\rightarrow\gamma\gamma}&\ =\ \frac{G_{\rm F}\alpha_{\rm em}^2M_\phi^3}{128\sqrt{2}\pi^3}\left|\frac{\lambda^\prime v }{6M_{S_1}^2}\mathcal{A}_0\left(\frac{M_\phi^2}{4M_{S_1}^2}\right)\right|^2
 \label{eq:diphoton}.
\end{align}
The Feynman diagrams for the $\phi\to\gamma\gamma$ process will be similar to those in Figs.~\ref{fig:glu_a} and~\ref{fig:glu_b}, with the gluons replaced by two photons and the $\alpha_{s}$ coupling  
substituted for the $\alpha_{\rm em}$ coupling. As earlier, we can now express the cross sections in these modes in terms of the partial widths. In the $gg$ channel,
\begin{align}
 \hat{\sigma}(gg\to \phi)=\frac{\pi^2 M_\phi}{8\hat s}\Gamma_{h\to gg}\delta(\hat{s}-M_\phi^2), 
 \end{align}
and in the $q\bar q$ channel,
\begin{equation}
\hat{\sigma} (q\bar{q}\to \phi)=\frac{4\pi^2M_\phi}{9\hat s}\Gamma_{\phi\to q\bar q}\delta(\hat{s}-M_\phi^2).
\end{equation} 
 The total width for $\phi$ can be expressed as, 
\begin{align}
\Gamma_\phi=&\left(\sum_{q=d,s,b}\Gamma_{\phi\rightarrow q\bar{q}}+\Gamma_{\phi\rightarrow gg}+\Gamma_{\phi\rightarrow\gamma\gamma}\right).
\label{eq:decaysinglet}
\end{align} 

We now present our numerical results. We begin with Fig.~\ref{fig:sing_Br}, where we show the variation of BRs of different decay modes of $\phi$. For the most part, the plots for the quarks overlap, as $\Gamma_{\phi\rightarrow q \bar q}$ is essentially independent of m$_{q}$ [see Eq.~\eqref{eq:singlet-phi-gg}].
Here, without any singlet-doublet mixing, $\phi$ can  decay only to down-type quarks or gluon or photon pairs. As a result, when $M_{S_1}$ increases, BR($\phi\to gg/\gamma\gamma$) decreases and BR($\phi\to q\bar q$) goes up if $M_{\nu_R}$ is held fixed. 
We see that for a $2$ TeV $S_1$, $\phi\to q \bar q$ is the dominant decay mode for $g^2y_\nu=1$, $M_{\nu_R}=1$ TeV (the BRs are independent of $\lambda^\prime$). 

In Figs.~\ref{fig:sing_BrCsa} and ~\ref{fig:sing_BrCsb}, we plot the scattering 
cross sections of $\phi$ in different decay modes at the $14$ TeV 
LHC, considering both the gluon and quark fusion processes. We show the production cross section times the branching 
ratio for all the modes, against $M_{S_1}$ and $M_\phi$. Note that,
in the parameter space that we consider, we find $\Gamma_\phi \ll M_\phi$, which makes the narrow width approximation used in our computation a valid one. Here, we use the same set of PDFs as in the $h_{125}$ case. To have some intuition about the strengths of different production channels, we scale the cross sections by $\sigma(gg\to h_{M_{\phi}})$, 
where $h_{M_\phi}$ represents a BSM Higgs 
whose couplings with the SM particles are the same as
those of $h_{125}$. Its production cross section in the gluon fusion mode can be computed from Eq.~\eqref{eq:gghprod}
after taking $M_{S_1}\to\infty$ in Eq.~\eqref{eq:gammaSM_h}, as
\begin{align}
 \hat{\sigma}(gg\to h_{M_\phi})\simeq\frac{G_{\rm F}\alpha_S^2M_{\phi}^4}{512\sqrt{2}\pi \hat{s}}\left|\mathcal{A}_{1/2}\left(\frac{M_\phi^2}{4m_t^2}\right)\right|^2\delta(\hat s-M_\phi^2).
\end{align}
Then we define the scaled cross sections as
 \begin{align}
 R_\phi (ii\to \phi) = \frac{\sigma(ii\to\phi)}{\sigma(gg\to h_{M_\phi})}.
 \label{eq:redsigma}
 \end{align}
 
In Figs.~\ref{fig:sing_Redgluona} and~\ref{fig:sing_Redgluonb}, we show the variation of $R_\phi$ with $M_{S_1}$ and $M_\phi$. Recall that a SM singlet $\phi$ cannot be
 produced at tree level. The leading order contribution to
$\sigma(ii\to\phi)$ starts at the one-loop level. In Fig.~\ref{fig:sing_Redgluonb}, we observe a crossover where the qqF becomes the dominant process over the 
ggF, i.e., 
$\sigma(q \bar{q}\to\phi) > \sigma(gg\to\phi)$ for a fixed value of LQ mass ($=2$ TeV). 
This is not a generic pattern and can be understood from Eqs.~\eqref{eq:singlet-phi-gg} 
and~\eqref{singlet-phi-gg} by varying a few
of the free parameters. For example, for 
a relatively large value of LQ mass ($M_{S_1}\geq 2$~TeV), one may obtain
$\Gamma_{\phi\rightarrow gg} \le \Gamma_{\phi\rightarrow q \bar{q}}$ when 
$\phi$ is not large, i.e., $M_\phi \le 250$~GeV. In this case, the quark fusion process would have leading contributions. 
If one increases $M_{S_1}$ further, $\Gamma_{\phi\rightarrow gg}$ decreases more rapidly than $\Gamma_{\phi\rightarrow q\bar q}$, with $M_\phi$ ensuring that the $q \bar{q}\to\phi$ process remains the dominant one for a larger range of $M_\phi$. For
example, if one sets $M_{S_1} \sim 3$~TeV, we find that quark fusion becomes 
dominant for $M_\phi \le 350$~GeV. However, the relative contributions are insensitive to the
value of $\lambda'$ chosen.

\subsection{Prospects at the LHC}
\noindent
It is clear that the scalar $\phi$ in our model would offer some novel and interesting phenomenology at the LHC. However, a detailed analysis is beyond the scope of this paper. Instead we now simply make a few comments on its prospects.

It may be possible to put a bound on $\sigma_\phi(M_\phi)$ from the dijet resonance searches. For example, the one performed 
by the CMS Collaboration at the $13$ TeV LHC~\cite{Sirunyan:2016iap} indicates that $\sigma_\phi\times$BR$(\phi\to gg)$ has to be less than about $1$ pb for $M_\phi=1$ TeV and about $20$ pb for $M_\phi=600$ GeV. Similarly, in the quark mode, $\sigma_\phi\times$BR$(\phi\to d\bar d+s\bar s+b\bar b)$ is less than about $1$ pb for $M_\phi=1$ TeV and about $5$ pb for $M_\phi=600$ GeV. Figure~\ref{fig:sing_BrCsb} (which is obtained for the $14$ TeV LHC) indicates that our choice of parameters easily satisfies this limit. Future searches in this channel would put stronger bounds on $\sigma_\phi$ and/or $M_\phi$.
The LHC has also searched for such a state in the $\gamma\gamma$ final states, though the present bound from this channel is weaker~\cite{Aaboud:2017yyg} than the dijet one. In our model, this channel is not at all promising, as can be seen in Figs.~\ref{fig:sing_BrCs} and~\ref{fig:sing_Redgluon}. Even the
HL-LHC might not be able to probe the singlet state in the $\gamma\gamma$ mode.

\section{Conclusion} \label{sec:conclu}
\noindent 
In this paper, we have considered a simple extension to the SM, in which we have a scalar LQ ($S_1$) with electromagnetic charge $1/3$ and heavy right chiral neutrinos. While the presence of both BSM particles may have its origin in a grand unified framework, we have simply considered their interactions at the TeV scale. The motivation for considering such an extension comes from the fact that it can accommodate Yukawa couplings of the down-type quarks that are enhanced compared to SM expectations.

We have shown that the LQ and the right chiral neutrinos can enhance the production cross section of the SM-like Higgs through a triangle loop. We have calculated the one-loop contributions to the Yukawa couplings of 
the down-type quarks. We have found the enhancements (which we have parametrized by the usual $\kappa_{d,s,b}$) 
for order $1$ new couplings and TeV scale new particles. 
We have then further extended our analysis to include a SM-singlet scalar $\phi$ in the model with a dimension-$1$ coupling with $S_1$ but no tree-level mixing with the SM-like Higgs.
We have found that, for a similar choice of parameters, the gluon fusion (through a LQ in the loop) and the quark fusion (mediated by a LQ and neutrinos in a loop) processes can lead to a significant cross section to produce $\phi$ at the LHC. They also enhance the decay width of the singlet. Interestingly, we have found that for a light
$\phi$, the quark fusion can become more important than the gluon fusion process as long as the mass of the LQ remains high ($\sim$ TeV).
In both cases, precise 
measurements of branching fractions or partial widths of the $125$ GeV SM-like Higgs or the singlet scalar, i.e., $h_{125}, \phi \to d \bar{d}, s \bar{s}, b \bar{b}$, would be crucial for testing or constraining the model at the high luminosity run of the LHC.

\begin{acknowledgments}\noindent
Our computations were supported in part by SAMKHYA: the High Performance Computing Facility provided by the Institute of Physics (IoP), Bhubaneswar, India. A. B. and S. M. acknowledge support from the Science and Engineering Research Board (SERB), DST, India under Grant No. ECR/2017/000517. We thank P. Agrawal for the helpful discussion. S. M. also acknowledges the local hospitality at IoP, Bhubaneswar during the meeting IMHEP-19, where this work was initiated. 
\end{acknowledgments}
\vfill
\bigskip
\bibliographystyle{JHEPCust.bst}
\bibliography{lepto_quark}

\end{document}